\documentclass[useAMS,usenatbib]{mn2e}
\pdfoutput=1
\usepackage{aas_macros}
\usepackage{epsfig}
\usepackage{amsmath}
\usepackage{amssymb}
\usepackage{leftidx}
\usepackage{natbib}
\usepackage{hyperref}
\usepackage{xcolor}
\def\beq{\begin{equation}}
\def\eeq{\end{equation}}
\def\ba{\begin{eqnarray}}
\def\ea{\end{eqnarray}}
\def\bal{\begin{align}}
\def\eal{\end{align}}
\def\bxi{{\mbox{\boldmath $\xi$}}}
\def\bnab{{\mbox{\boldmath $\nabla$}}}

\def\icarus{Icarus}
\def\cqg{Class.\ Quantum Grav.}

\begin{document}

\title[Supernova Seismology] {Supernova Seismology: Gravitational Wave Signatures of Rapidly Rotating Core Collapse}

\author[J. Fuller et al.]{Jim Fuller$^{1,2}$,\thanks{Email: jfuller@caltech.edu}
Hannah Klion$^1$,
Ernazar Abdikamalov$^{3}$, and
Christian D. Ott$^1$\\
\\$^1$TAPIR, Walter Burke Institute for Theoretical Physics, Mailcode 350-17, California Institute of Technology, Pasadena, CA 91125, USA
\\$^2$Kavli Institute for Theoretical Physics, Kohn Hall, University of California, Santa Barbara, CA 93106, USA
\\$^3$Physics Department, School of Science and Technology, Nazarbayev University, 53 Kabanbay Batyr Ave., Astana, 010000, Kazakhstan
}

\label{firstpage}
\maketitle

\begin{abstract}

Gravitational waves (GW) generated during a core-collapse supernova open a window into the heart of the explosion. At core bounce, progenitors with rapid core rotation rates exhibit a characteristic GW signal which can be used to constrain the properties of the core of the progenitor star. We investigate the dynamics of rapidly rotating core collapse, focusing on hydrodynamic waves generated by the core bounce and the GW spectrum they produce. The centrifugal distortion of the rapidly rotating proto-neutron star (PNS) leads to the generation of axisymmetric quadrupolar oscillations within the PNS and surrounding envelope. Using linear perturbation theory, we estimate the frequencies, amplitudes, damping times, and GW spectra of the oscillations. Our analysis provides a qualitative explanation for several features of the GW spectrum and shows reasonable agreement with nonlinear hydrodynamic simulations, although a few discrepancies due to non-linear/rotational effects are evident. The dominant early postbounce GW signal is produced by the fundamental quadrupolar oscillation mode of the PNS, at a frequency $0.70 \, {\rm kHz} \lesssim f \lesssim 0.80\,{\rm kHz}$, whose energy is largely trapped within the PNS and leaks out on a $\sim\!10$ ms timescale. Quasi-radial oscillations are not trapped within the PNS and quickly propagate outwards until they steepen into shocks. Both the PNS structure and Coriolis/centrifugal forces have a strong impact on the GW spectrum, and a detection of the GW signal can therefore be used to constrain progenitor properties.

\end{abstract}


\section{Introduction}
\label{intro}

Rotating iron core collapse in a massive star ($M\gtrsim 8\,M_\odot$,
resulting in a core-collapse supernova [CC SN]) was one of the first
potential sources of gravitational waves (GWs) considered in the
literature (\citealt{weber:66,ruffini:71}; see \citealt{ott:09} for a
historial overview). GWs are of lowest-order quadrupole waves and
rotation naturally drives quadrupole deformation (oblateness) of the
homologously ($v \propto r$) collapsing inner core of a rotating
massive star. When the inner core reaches nuclear densities, the
nuclear equation of state stiffens, stopping the collapse of the inner
core. The latter overshoots its new equilibrium, bounces back (a process called ``core
bounce'') and launches the hydrodynamic supernova shock at its
interface with the still collapsing outer core. Subsequently, the
inner core rings down, shedding its remaining kinetic energy in a few
pulsations, then settles to its new postbounce equilibrium and becomes
the core of the newly formed proto-neutron star (PNS). The entire
rotating bounce--ring down process involves rapid changes of the inner
core's quadrupole moment and tremendous accelerations. The resulting
GW burst signal has been investigated extensively both with
ellipsoidal models (e.g., \citealt{saenzshapiro:78}) and with detailed
multi-dimensional numerical simulations
(e.g., \citealt{mueller:82,moenchmeyer:91,yamadasato:95,zwerger:97,dimmelmeier:02,kotake:03,ott:04,dimmelmeier:08,obergaulinger:06b,ott:07prl,ott:12a}). 

Based on this extensive volume of work, it is now clear that rotating core
collapse proceeds mostly axisymmetrically and that nonaxisymmetric
dynamics sets in only within tens of milliseconds after bounce
\citep{ott:07prl,scheidegger:08,scheidegger:10,kuroda:14}. 
%Hence, the resulting GW signal is linearly polarized and beamed away from the axis of rotation.
Only rapidly rotating iron cores (producing PNSs
with central spin periods $\lesssim$ $5\,\mathrm{ms}$) generate
sufficiently strong GW signals from core bounce to be detected
throughout the Milky Way by Advanced-LIGO-class GW observatories
(\citealt{aligo,ott:12a}, hereafter O12). Since the cores of most massive stars are
believed to be slowly rotating at core collapse ($\gtrsim 90\%$,
e.g., \citealt{heger:05,ott:06spin,langer:12}), the detection of GWs
of a rotating core collapse event may be exceedingly rare and GW
emission in CC SNe may be dominated by neutrino-driven convection
instead \citep[e.g.,][]{mueller:97,mueller:04,ott:09,kotake:13review,
  ott:13a,mueller:e12,mueller:13gw,murphy:09}. However, if a rapidly
rotating core collapse event were to be detected, it could possibly be
linked to an energetic CC SN driven by magnetorotational coupling
\citep[e.g.,][]{burrows:07b,takiwaki:12,moesta:14b}. \cite{abdikamalov:14} (A14 hereafter) have shown that the angular
momentum of the inner core can be measured from the observed GW signal.

\begin{figure}
\begin{center}
\includegraphics[scale=0.45]{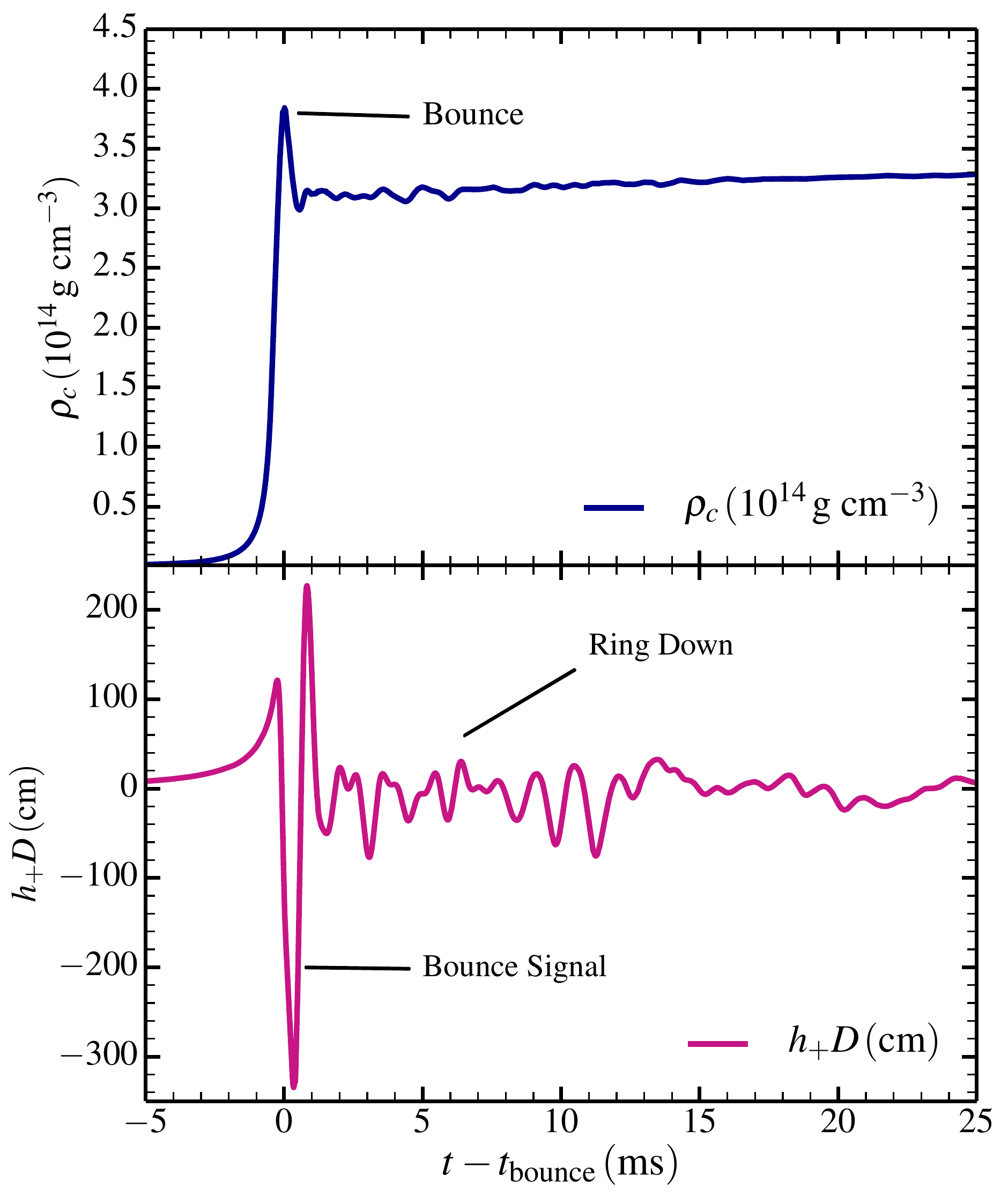}
\end{center} 
\caption{ \label{NSstrain} {\bf Top:} Central density $\rho_c$ of the rapidly rotating A3O05 simulation of A14. Time is measured from the moment of bounce, defined as the moment when $\rho_c$ peaks. {\bf Bottom:} GW strain from the same simulation, computed at a distance $D=10\,{\rm kpc}$. The large amplitude signal near bounce is the ``bounce signal" created primarily by the quasi-radial bounce of the centrifugally distorted inner core (O12, A14). The smaller amplitude ``ring down" oscillations after bounce are created primarily by quadrupolar oscillations of the PNS.}
\end{figure}

The morphology of the GW signal from rotating core collapse, bounce,
and ring down is uniform across the entire parameter space of
plausible initial conditions (\citealt{dimmelmeier:07,dimmelmeier:08}, O12) and essentially 
independent of progenitor star mass. It consists of a first prominent peak associated with core
bounce (the ``bounce signal'', cf.~Figure~\ref{NSstrain}) and a
lower-amplitude but longer duration oscillatory ring down signal, persisting
for $\sim$ $20\,\mathrm{ms}$ after bounce. The ring down signal is
peaked at a GW frequency of $0.7-0.8\,\mathrm{kHz}$ (somewhat dependent
on equation of state and rotation rate, \citealt{dimmelmeier:08}; O12;
\citealt{klion:15}, in prep, hereafter K15) and may be correlated with variations in the early postbounce
neutrino luminosity (O12), suggesting that the ring-down
oscillations of the PNS (here defined as the inner $20 \, {\rm km}$ of the postbounce star) are connected to the excitation of a global
PNS oscillation mode at core bounce (O12).

While the GW signal of rapidly rotating CC SNe can be
straightforwardly computed from complex nonlinear multi-dimensional
hydrodynamic simulations, its features are not understood at a
fundamental level. The goal of the present investigation is to provide such a basic understanding of the signal features. To do this, we employ semi-analytical, linear calculations of the wave-like fluid response produced by the bounce of the inner core. These calculations shed light on the physical mechanisms responsible for the GW signals discussed above, and their simplicity complements the complexity of the simulation results. However, our methods only provide a qualitative explanation for GW signals, the simulations are needed for precise quantitative predictions. Although oscillations of PNSs have previously been examined in several papers (see, e.g., \citealt{ferrari:03,ferrari:04}), these works focus on PNS oscillation modes well after ($\gtrsim 100 \, {\rm ms}$) bounce, and they do not investigate the physics of waves excited by the bounce itself.

We find that the postbounce fluid response has a few distinguishing
characteristics. First, the bounce excites a train of radial,
outwardly propagating acoustic waves. Because the background structure
is centrifugally distorted by rotation, these waves are only
quasi-radial and contain a quadrupole moment, allowing them to emit
the GW responsible for the bounce signal discussed above. Second, the
centrifugal distortion of the progenitor leads to the excitation of a
train of axisymmetric quadrupolar waves. Some of these waves are
reflected at the edge of the PNS, causing them to interfere to create
standing waves, whose energy is peaked near the oscillation ``modes"
of the PNS. The dominant GW signal is produced by the fundamental PNS
oscillation mode at a frequency $f_{\rm f-mode} \sim 0.75\,{\rm kHz}$,
accounting for the most prominent peak in the GW ring down signal. The
GW signal damps on $10\,{\rm ms}$ timescales as the wave energy leaks
out of the PNS into the surrounding envelope.

Our paper is organized as follows. In Section \ref{oscillations} we
introduce our semi-analytical framework for calculating the spectrum
of waves excited at core bounce, and we discuss the properties of the
resulting waves. Section \ref{nonad} investigates the subsequent wave
damping, and discusses the complications introduced
rotational and relativistic effects. In Section \ref{GW} we present the
GW spectra produced by the waves and compare with the GW spectra seen
in simulations. We conclude in Section \ref{disc} with a discussion of
our results and future avenues for theoretical development.

\section{Oscillations Excited at Bounce}
\label{oscillations}

As described above, the GW spectrum of a rapidly rotating supernova near core bounce consists of a bounce signal and a ring down signal. The bounce signal has a short duration in time and is thus characterized by a broad frequency spectrum, while the ring down signal lasts longer and has a spectrum peaked at discrete frequencies. Our main goal is to understand the physics of the ring down signal, although our methods also shed some light on the spectrum of the bounce signal.

During CC, the inner core of a massive star progenitor collapses into a PNS, while the outer core forms the shocked region surrounding the PNS core during the on-going supernova. In this paper, we refer to the inner $r \lesssim 30\,{\rm km}$ ($\rho \gtrsim 10^{12}\,{\rm g}\,{\rm cm}^{-3}$) as the PNS, while the low density surrounding regions are the envelope. We choose this definition because the bounce excited waves can become trapped in the inner $\sim \! 30 \,{\rm km}$. However, note that at a few ms after bounce, the PNS only has a mass of $M_{\rm PNS} \! \sim \! 0.6 M_{\odot}$ while the mass within the inner $300\,{\rm km}$ is $M\!\sim\!1 M_\odot$, and we therefore expect all of the material within our computational domain to eventually accrete onto the central compact object. In the brief (less than a second) period following the PNS bounce, but preceding the supernova, the inner $\sim \!\! 100$ km (i.e., regions below the shock radius) of the supernova core is in approximate hydrostatic equilibrium \citep{janka:01}.

\subsection{Models}
\label{model}

\begin{figure}
\begin{center}
\includegraphics[scale=0.45]{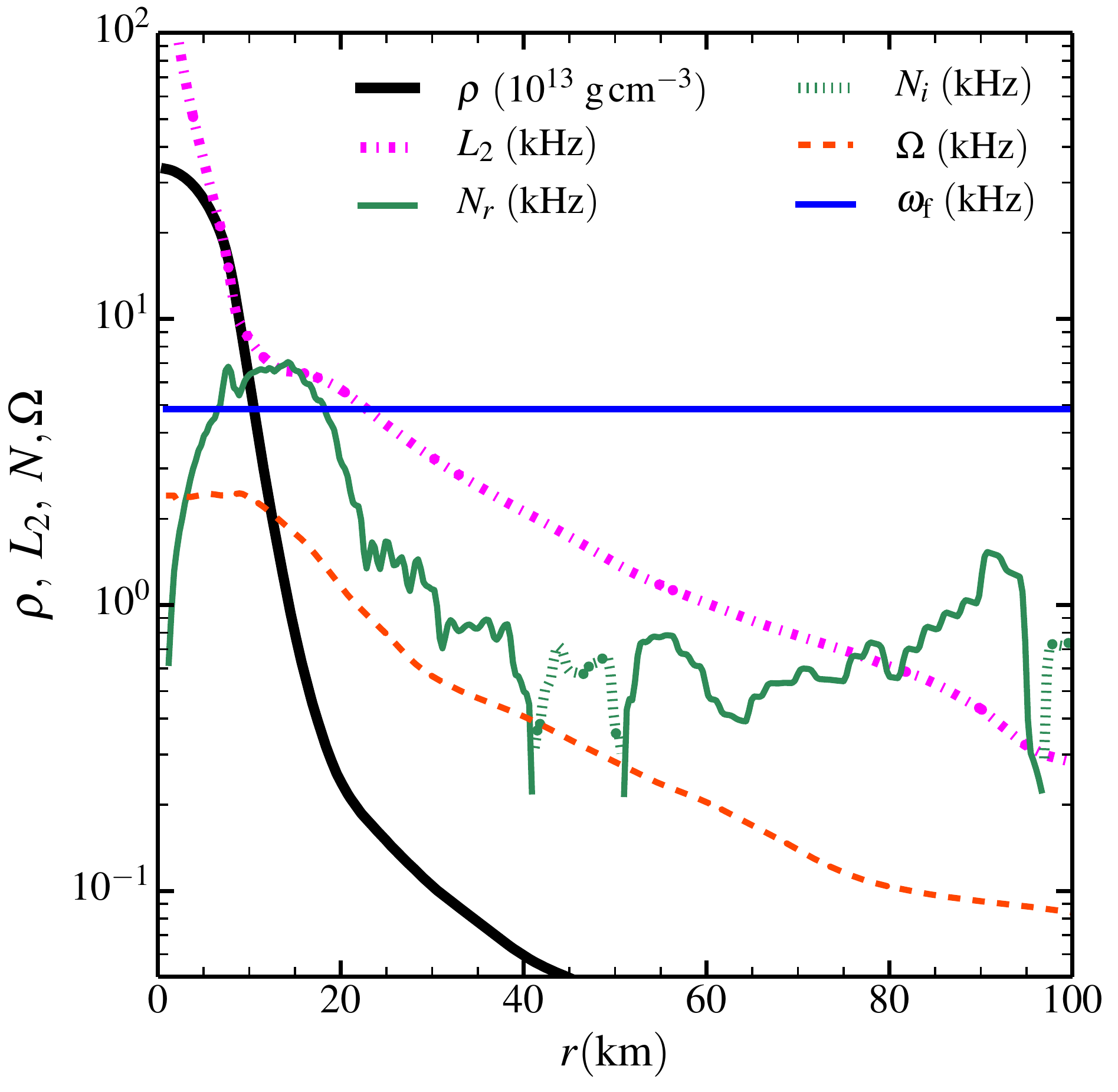}
\end{center} 
\caption{ \label{NSstruc} Density ($\rho$), Brunt-V\"{a}is\"{a}l\"{a} frequency ($N$), Lamb frequency ($L_2$), and angular spin frequency ($\Omega$) profiles in the inner 100 km of the early postbounce evolution of model A3O04 of A14. Positive values of $N^2$ are denoted by the real part of the Brunt-V\"{a}is\"{a}l\"{a} frequency ($N_r$) while negative values of $N^2$ correspond to imaginary values of the Brunt-V\"{a}is\"{a}l\"{a} frequency ($N_i$). The PNS occupies the inner $\sim 30$ km of the remnant. Waves near the f-mode frequency have $f \sim 0.8 \,{\rm kHz}$, corresponding to angular frequencies $\omega = 2 \pi f \sim 5$ kHz, marked by the horizontal line. The model is made by averaging the simulation output over 10 ms, starting 3 ms after bounce. At these times, the shock is located near $r\sim 95$ km.}
\end{figure}

To generate background models for wave excitation and propagation within the postbounce supernova structure, we use simulation outputs generated by A14. These simulations are run with the 2-dimensional version of the CoCoNuT code (\citealt{dimmelmeier:02,dimmelmeier:05}) in axisymmetry and conformally flat general relativity. These approximations are all appropriate for understanding the axisymmetric waves of interest. In the postbounce phase, a neutrino leakage/heating scheme approximates the effects of neutrinos. We choose snapshots of the supernova structure beginning 3 ms after bounce and average the next 10 ms of evolution (sampled by snapshots every 1 ms) to determine the background structure. This procedure smooths out most of the waves, turbulence, and other short-lived features present within the supernova core without allowing for significant evolution of the background structure. Our fiducial model is the A3O04 model of A14. The rapid rotation of this model is sufficient to generate a rotationally dominated GW signal, but slow enough to be reasonably approximated by our semi-analytical techniques described below.

Figure \ref{NSstruc} shows a density profile and propagation diagram for the central 100 km of our fiducial model, averaged over the time period 3-12 ms after bounce. The inner $\sim \! 30\,{\rm km}$ of the star comprise the high density PNS, which is surrounded by a much lower density envelope. The shock radius is near $r \sim 95$ km during this time, and the shock does not strongly affect the nature of waves propagating near the PNS. In our models, the PNS is always stably stratified; convection driven by a lepton gradient does not develop until later times, which we do not study here.

\subsection{Wave Excitation and Computation}

The sudden deceleration at core bounce excites waves which propagate within the PNS and surrounding material. Here, we semi-analytically calculate the spectrum of waves excited by the bounce. We use linear and adiabatic approximations for displacements from the background state, which we assume to be in hydrostatic equilibrium. We discuss non-adiabatic and non-linear effects in Section \ref{nonad}. We also temporarily ignore special/general relativistic effects and the impact of Coriolis and centrifugal forces, which we address in Section \ref{rotation}

Applying the approximations listed above, the linearized momentum equation is
\beq
\label{momeq}
\frac{\partial^2}{\partial t^2} \bxi = -\frac{1}{\rho} \bnab \delta P - \bnab \delta \Phi - g \frac{\delta \rho}{\rho} {\hat{\bf r}} + {\bf f}({\bf r},t)\,\,.
\eeq
Here, $\bxi$ is the Lagrangian displacement, $\rho$ is the density, $g$ is the gravitational acceleration, $\delta P$ and $\delta \rho$ are the Eulerian pressure and density perturbations, and ${\bf f}$ is the force per unit mass provided by the bounce. We also use the continuity equation 
\beq
\label{cont}
\delta \rho + \bnab \cdot \big( \rho \bxi \big) = 0\,\,,
\eeq
the adiabatic equation of state
\beq
\label{drho}
\delta \rho = \frac{1}{c_s^2} \delta P + \frac{\rho N^2}{g} \xi_r\,\,,
\eeq
and Poisson's equation
\beq
\label{poisson}
\nabla^2 \delta \Phi = 4 \pi G \delta \rho\,\,.
\eeq
Here, $\delta \Phi$ is the gravitational potential perturbation, $c_s$ is the sound speed, $N$ is the Brunt-V\"{a}is\"{a}l\"{a} frequency, and $G$ is Newton's gravitational constant. 

Waves are excited by the force applied during the bounce of the inner core. For a spherically symmetric collapse, the strength of the force will be comparable to the amount of force required to halt the collapse of a shell of material at radius $r$ infalling at the escape velocity. Its direction will 
be radially outwards. A rough estimate of the magnitude of the force per unit mass is
\beq
\label{f1}
{\bf f}(r) \sim g {\hat{\bf r}}\,\,,
\eeq
where $g$ is the local gravitational acceleration in the hydrostatic postbounce material, and ${\hat{\bf r}}$ is the radial unit vector. The force peaks at the moment of the bounce, which we define as $t=0$. The duration of the bounce is approximately equal to the local dynamical time, $t_{\rm bounce} \sim t_{\rm dyn} = \sqrt{r^3/GM(r)}$. We approximate the time dependence of the bounce as a Gaussian of width equal to the dynamical time such that 
\beq
\label{f2}
{\bf f}(r,t) \sim g e^{-(t/t_{\rm dyn})^2} {\hat{\bf r}}\,\,.
\eeq
Note that both $g$ and $t_{\rm dyn}$ are functions of radius, so both the magnitude and duration of the force are strongly dependent on radial position. Because the gravitational acceleration $g$ peaks in the core of the PNS, the forcing is concentrated in this region (see Figure \ref{NSwavefunction}). 

In a non-rotating progenitor, the collapse would be spherically symmetric and only radial oscillations would be generated. In rapidly rotating progenitors, the background structure is centrifugally distorted, leading to the generation of axisymmetric non-radial waves. We may expect the degree of the non-spherical component of the force to be proportional to the centrifugal distortion of the collapsing star, $\epsilon$, defined such that the background density structure has the form
\beq
\label{isodens}
\rho(r,\theta) = \rho(r) \big[1 + \epsilon(r) \, Y_{20}(\theta) \big].
\eeq
Here, $Y_{20}$ is the $l=2$, $m=0$ spherical harmonic, and $\rho(r)$ is the spherically averaged density profile. The centrifugal acceleration has magnitude $\sim r \Omega^2$, where $\Omega$ is the local spin frequency. Thus, the centrifugal distortion scales (in the limit $\epsilon \ll 1$) as
\beq
\epsilon \sim (\Omega/\Omega_{\rm dyn})^2\, , 
\eeq
where $\Omega_{\rm dyn} = t_{\rm dyn}^{-1}$ is the local dynamical frequency. In Appendix \ref{equations}, we show that the perturbation in the bounce force per unit mass due to the centrifugal distortion is
\begin{align}
\label{force}
{\rm {\bf  \delta f} } \simeq A \sqrt{\frac{2}{\pi}} \, \epsilon \, g \, e^{-(t/t_{\rm dyn})^2} \bigg[ 2 \, Y_{20} \, {\hat {\bf r} } + r \bnab_\perp Y_{20} \bigg] \,\,,
\end{align}
where $\bnab_\perp$ is the non-radial component of the gradient, and $A \! \sim \! 1$ parameterizes the magnitude of the force. 

With an estimate of the forcing function in hand, we can solve Equations \ref{momeq}-\ref{poisson} for the forced response of the fluid due to the bounce. It is most convenient to solve these equations in the frequency domain rather than the time domain. To do this, we decompose all perturbation variables into their components per unit frequency, e.g.,
\beq
\label{xifreq}
\bxi(t) = \int d \omega' \bxi_{\omega'}(\omega') e^{-i\omega' t}\,\,.
\eeq
Inserting this expression into Equation \ref{momeq}, multiplying by $e^{i \omega t}$, and integrating over time, we obtain
\beq
\label{momfreq}
-\omega^2 \bxi_\omega + \frac{1}{\rho} \bnab \delta P_\omega + \bnab \delta \Phi_\omega + g \frac{\delta \rho_\omega}{\rho} {\hat{\bf r}} = {\rm {\bf  \delta f} }_\omega \,\,,
\eeq
with 
\beq
\label{fom}
{\rm {\bf  \delta f} }_\omega =  \delta {\rm f}_\omega \bigg[ 2 \, Y_{20} \, {\hat {\bf r} } + r \bnab_\perp Y_{20} \bigg] \, ,
\eeq
and
\beq
\label{fval}
 \delta {\rm f}_\omega \equiv \frac{A}{\sqrt{2} \pi} \, \epsilon \, g \, t_{\rm dyn} \, e^{-(\omega t_{\rm dyn}/2)^2} .
\eeq
In Equation \ref{momfreq}, each perturbed quantity is the perturbation per unit frequency. Similar equations can easily be derived from Equations \ref{cont}-\ref{poisson}, and are given in Appendix \ref{equations}. The Gaussian frequency dependence of the forcing term in Equation \ref{momfreq} entails that only waves with angular frequencies $\omega \lesssim 1/t_{\rm dyn}$ will be strongly excited at a particular location. The dynamical time is smallest at the center of the PNS where the density is highest, we therefore do not expect significant excitation of waves with angular frequencies $\omega \gg \sqrt{G \rho_c} \sim 10\,{\rm kHz}$. 

To solve for the frequency component response, we must also implement boundary conditions. At the center of the PNS we adopt the standard central reflective boundary conditions (see Appendix \ref{equations}). However, in the outer regions of our computational domain ($r \sim 250$ km), there is no surface at which the waves will reflect.\footnote{A possible exception is the shock front, however, we expect the waves to be largely dissipated before they reach the shock (see Section \ref{nonad}).} Instead we expect the waves to propagate outwards until they dissipate (see Section \ref{nonad}). In the outer regions, the waves generally behave like acoustic (pressure) waves because their frequencies are greater than the local dynamical frequency. We therefore adopt a radiative outer boundary condition (described in Appendix \ref{equations}) that ensures only outgoing acoustic waves exist at the outer boundary.

\subsection{Wave Propagation and Characteristics}
\label{waveprop}

The behavior of waves can be understood from their WKB dispersion relation
\beq
\label{disp}
k_r^2 = \frac{(L_l^2-\omega^2)(N^2-\omega^2)}{\omega^2 c_s^2}\,\,,
\eeq
where $k_r$ is the radial wavenumber, 
\beq
\label{lamb}
L_l^2 = \frac{l(l+1)c_s^2}{r^2}
\eeq
is the Lamb frequency squared, and $N^2$ is the Brunt-V\"{a}is\"{a}l\"{a} frequency squared. In regions where $\omega > L_l$ and $\omega>N$, waves behave like acoustic waves, while they behave like buoyancy (gravity) waves (not to be confused with GWs) where $\omega < L_l$ and $\omega < N$. Radial profiles of $N$ and the quadrupolar Lamb frequency, $L_2$, are shown in Figure \ref{NSstruc}. Recall that typical bounce-excited waves have $l=0$ or $l=2$ and angular frequencies near $\omega \! \sim \! \sqrt{G \rho_c} \! \sim \! 5 \, {\rm kHz}$, corresponding to $f \! \sim 0.8 \, {\rm kHz}$.

\begin{figure*}
\begin{center}
\includegraphics[scale=0.6]{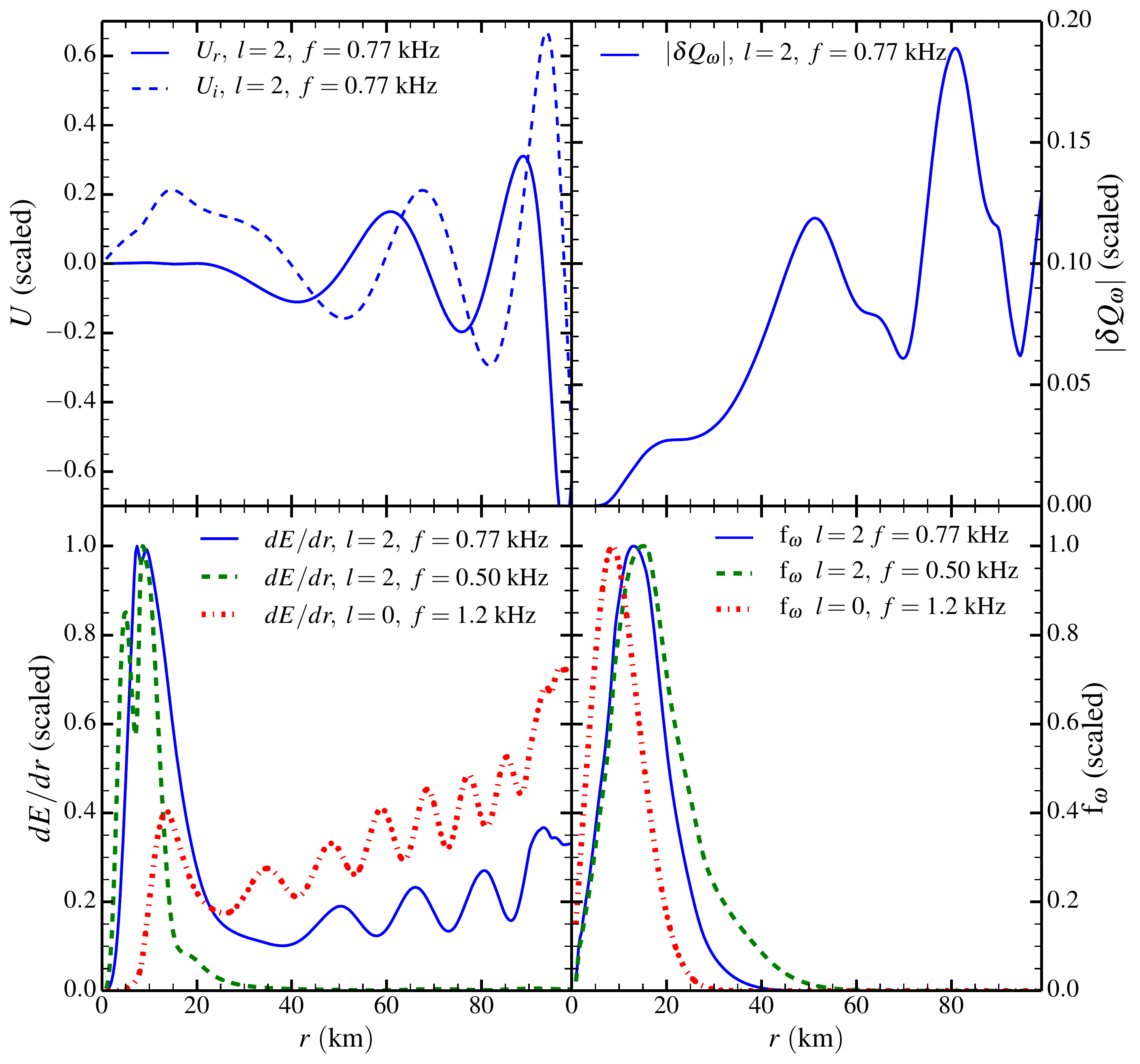}
\end{center} 
\caption{ \label{NSwavefunction} {\bf Top Left:} Real ($U_r$) and imaginary ($U_i$) parts of the radial wave displacement per unit frequency (Equation \ref{xir}) for bounce-exited waves with $f=0.77$ kHz. {\bf Top Right:} Magnitude of the quadrupole moment $|\delta Q_\omega|$ (Equation \ref{quad2}) for the same waves. {\bf Bottom Left:} Time integrated wave energy per unit radius (Equation \ref{edens2}) for waves of different frequencies. The three shown frequencies approximately correspond to the PNS quadrupolar ($l=2$) f-mode ($f=0.77$ kHz), the PNS $l=2$ g$_1$-mode ($f=0.50$ kHz), and an outgoing quasi-radial wave ($f=1.2$ kHz). {\bf Bottom Right:} Strength of the excitation force ${\rm f}_\omega$ (Equation \ref{fval}). All quantities are normalized via their maximum values within the computational domain (see Figure \ref{NSnonlin} for dimensional values of the wave amplitude).  }
\end{figure*}

Near the outer boundary, $L_2$ and $N$  are much smaller than typical wave frequencies $\omega$, and in this limit the dispersion relation reduces to that of acoustic waves,
\beq
\label{dispsound}
k_r^2 \simeq \frac{\omega^2}{c_s^2}\,\,.
\eeq
The wavelength shortens as the waves propagate outwards into regions with smaller sound speeds. Moreover, the WKB wave amplitude scales as $\xi_r \propto (\rho c_s )^{-1/2}$, and so the waves grow in amplitude (although their energy flux remains constant) as they propagate outwards. These factors cause the waves to become increasingly non-linear as they propagate outwards, i.e., they steepen into shocks. The radial group velocity of the waves in the outer regions is simply $v_g \simeq c_s$, so that wave energy travels outwards at the sound speed. 

At the center of the PNS, $L_2 \rightarrow \infty$ and $N \rightarrow 0$, so quadrupolar waves are evanescent for $r \lesssim 5\,{\rm km}$. In this region, the response is not wave-like, but is characterized by the coherent fundamental mode-like oscillation of the PNS. The value of $N^2$ is large where the density gradient is large at radii $5 \, {\rm km} \lesssim r \lesssim 20 \, {\rm km}$. This region of the star can harbor buoyancy waves for frequencies $\omega \lesssim 5$ kHz.

Figure \ref{NSwavefunction} shows a plot of the radial component of the wave displacement per unit frequency, $U$, for a quadrupolar wave with angular frequency $\omega \simeq 5\,{\rm kHz}$ ($f \simeq 0.8$ kHz) as a function of radius. The wave contains both a real part ($U_r$) and an imaginary part ($U_i$), which are perpendicular in phase for a propagating wave and in phase for a standing wave. These waves reflect at the edge of the PNS (at radii of $r\sim 20 \,{\rm km}$), so the response within the PNS is composed of both ingoing and outgoing waves, which interfere to produce a standing wave, or oscillation mode. There are no nodes in the radial wave function within the PNS ($r\lesssim 15\,{\rm km}$), therefore we refer to this mode as the fundamental mode (f-mode) of the PNS.\footnote{In our models the f-mode has $\omega<N$ in much of the PNS, therefore it has gravity mode characteristics, which we discuss in more depth in Section \ref{rotation2}.} The standing waves are not totally reflected, and gradually leak into the surrounding material. For $r\gtrsim 50 \, {\rm km}$, the real and imaginary component of $U(r)$ are perpendicular in phase, characteristic of an outwardly propagating acoustic wave.

Figure \ref{NSwavefunction} also plots the strength of the forcing, ${\rm f}_\omega$ (integrand of Equation \ref{fval}), and the time-integrated wave energy per unit radius, given by Equation \ref{edens2}, for waves of different frequencies. The forcing is localized to near the PNS, especially for higher frequency waves. For quadrupolar waves, the displacements are largest outside the inner core $(r\gtrsim30\,{\rm km})$, although the wave energy density is primarily localized to the inner $\sim$20 km. This indicates that quadrupolar waves are trapped within the PNS, and only gradually leak out into the outer regions. For quasi-radial waves, the time-integrated wave energy is smaller in the PNS and larger in the envelope, indicating these waves are not well-trapped in the PNS and quickly propagate outward. We shall see in Section \ref{GW} that wave energy is sharply peaked at characteristic frequencies that correspond to the PNS oscillation mode frequencies. The waves shown in Figure \ref{NSwavefunction} have frequencies approximately corresponding to the PNS $l=2$ fundamental oscillation mode (f-mode, $f \simeq 0.8\,{\rm kHz}$), the first $l=2$ gravity mode (g$_1$-mode, $f\simeq 0.5\,{\rm kHz}$), and an outgoing $l=0$ pressure wave (p-wave, $f \simeq 1.2\,{\rm kHz}$).
     \!\!\footnote{We label the modes by the number of nodes in the radial displacement $U_r$ within the PNS. The f-mode has no nodes, while the $g_1$-mode has one node, and so on.} 

Finally, Figure \ref{NSwavefunction} shows the $m=0$ component of the wave quadrupole moment per unit frequency,
\beq
\label{quad1}
\delta Q_\omega = \int dV \ r^2 \delta \rho_\omega({\bf r}) Y_{20}^*\,\,.
\eeq
For the $l=2$, $m=0$ waves, Equation \ref{quad1} reduces to
\beq
\label{quad2}
\delta Q_\omega = \int dr \ r^4 \delta \rho_\omega(r)\,\,.
\eeq
The magnitude of the quadrupole moment is somewhat oscillatory, but generally increases with radius. In the absence of wave damping, GWs are more efficiently generated as the waves propagate outward, however, we find below that waves are generally dissipated before reaching large radii.

\begin{figure}
\begin{center}
\includegraphics[scale=0.45]{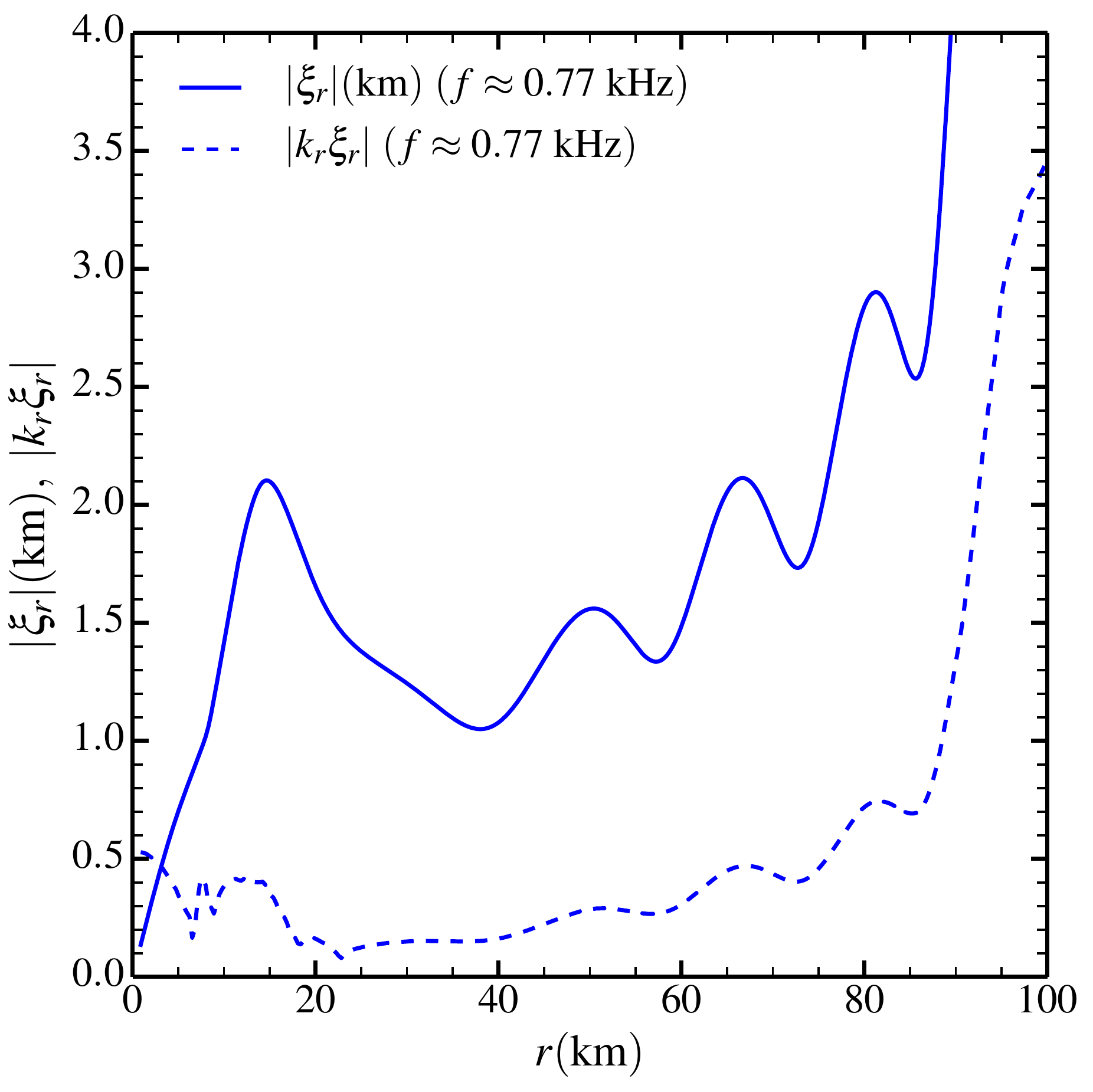}
\end{center} 
\caption{ \label{NSnonlin} Approximate maximum amplitudes of the radial displacements, $\xi_r$, associated with waves with frequencies near the $l=2$ f-mode. We also plot the approximate non-linearity parameter, $|k_r \xi_r|$, associated with the waves. Waves are highly non-linear and are expected to generate shocks where $|k_r \xi_r| \gtrsim 1$. Since the waves are somewhat non-linear, our linear results are only approximate, and may differ from simulation results by a factor of order unity. Wave damping is not included in this plot, and these amplitudes will likely be diminished at large radii and late times. }
\end{figure}

The approximate physical amplitude of the waves is shown in Figure \ref{NSnonlin}. Precise wave amplitudes require an integral over the response per unit frequency at a given time $t$. However, as discussed in Section \ref{GW}, the wave response is sharply peaked near discrete frequencies approximately corresponding to oscillation mode eigenfrequencies. We can therefore integrate the response over each of these narrow peaks to estimate wave amplitudes. We justify this procedure in Appendix \ref{nonlinear}. The resulting amplitudes imply radial wave displacements of $\sim $kilometers and velocities of  a few percent the speed of light. These amplitudes are moderately non-linear, which we discuss in more detail below.

\section{Wave Damping and Rotation}
\label{nonad}

\subsection{Wave Damping}

The analysis presented above did not include any sources of wave damping. In reality the waves will damp out on a relatively short ($\sim \! 10\,{\rm ms}$) timescale. Our goal here is to identify the primary damping mechanism and estimate the wave lifetime. 

Wave damping due to photon diffusion is orders of magnitude longer than any relevant timescales and can be ignored. The radiative diffusion of neutrinos, however, could potentially provide a significant damping mechanism. Indeed, the simulations of O12 show correlated neutrino and GW strain oscillations, implying that some wave energy may be carried away by neutrinos. However, these simulations also showed very little difference between the gravitational waveforms with neutrino leakage turned on or off. \cite{ferrari:03} find that neutrinos damp PNS oscillation modes on a neutrino diffusion time scale of $\sim \! {\rm seconds}$. We therefore consider it unlikely that neutrinos can significantly damp the waves considered here on timescales as short as tens of milliseconds. 

GWs generated by the waves carry away wave energy and could potentially be an important source of wave damping. As the waves propagate outward, their quadrupole moment increases (see Figure \ref{NSwavefunction}) and so their GW energy emission rate increases. However, the waves also become increasingly non-linear (see Figure \ref{NSnonlin}). We find that waves nearly always become non-linear before they radiate a significant fraction of their energy into GWs. This is consistent with simulations \citep{ott:09,kotake:13review}, which find that the energy radiated in GWs is a small fraction of the energy contained in fluid wave-like motions. Therefore it is unlikely that GW emission is a significant source of damping for most waves.
\!\! \footnote{High frequency waves ($f \gtrsim 2$ kHz) may radiate much of their energy in GWs due to the $f^6$ dependence of the GW energy emission rate. However, since these waves are weakly excited at bounce, they contain little energy and cannot generate a strong GW signature. }

If non-linear wave breaking occurs, the waves generate shocks at which point their energy is rapidly converted into increased entropy of the fluid where the shock forms. We expect the waves to non-linearly dissipate when their amplitude is comparable to their wavelength, i.e., when
\beq
\label{nonlin}
k_r \xi_r \sim 1\,\,.
\eeq
Figure \ref{NSnonlin} shows an estimate of the physical displacements of the waves, and their degree of non-linearity. The largest amplitude waves (with frequencies $f \sim 0.8\,{\rm kHz}$) are moderately non-linear, reaching amplitudes $k_r \xi_r \sim 0.5$ within the PNS. Indeed, the simulations of A14 and K15 show evidence for the first harmonic of these waves in the GW spectra,\footnote{These peaks are more prominent in postbounce spectra, i.e., spectra where the bounce is windowed out. This allows the non-linear harmonic peak to be separated from the broad spectrum of GWs contributing to the bounce signal.} which is one possible outcome of non-linear wave coupling. In the absence of other sources of damping, the waves become very non-linear when $r \gtrsim 90$ km, and will non-linearly break if they are able to propagate that far. We also find that g-modes are very non-linear within the PNS, and likely break and dissipate within the PNS on short timescales. 

\begin{figure}
\begin{center}
\includegraphics[scale=0.43]{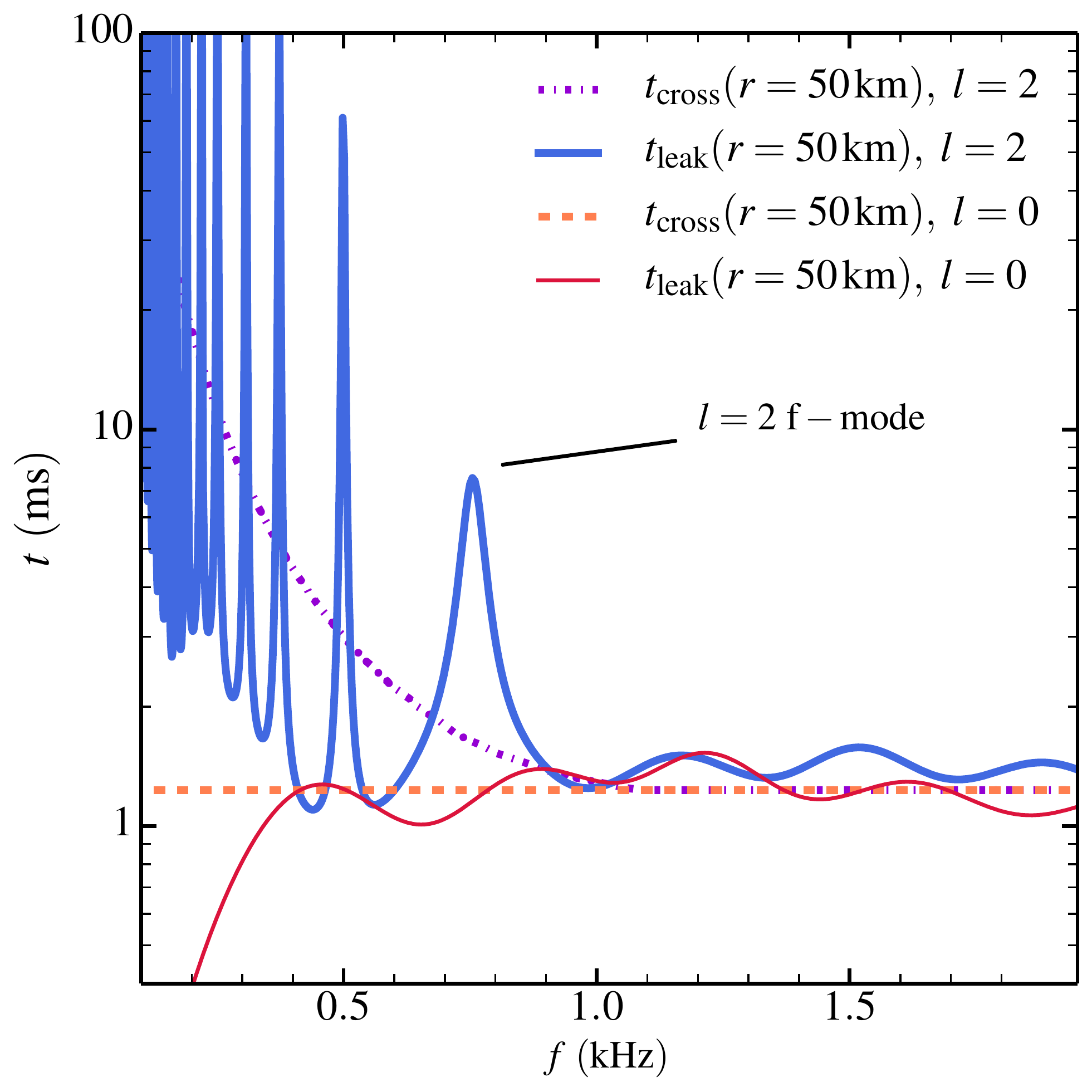}
\end{center} 
\caption{ \label{NStime} Wave crossing timescale $t_{\rm cross}$ and wave leakage timescale $t_{\rm leak}$ across the inner 50 km of the postbounce supernova structure as a function of wave frequency for $l\!=\!2$ and $l\!=\!0$ waves. Each peak in $t_{\rm leak}$ corresponds to a PNS oscillation mode; the peak at $f\sim 0.77\,{\rm kHz}$ is the $l\!=\!2$ f-mode. Quasi-radial ($l\!=\!0$) waves have $t_{\rm cross} \approx t_{\rm leak}$ because these waves propagate outward without significant reflection.}
\end{figure}

The arguments above suggest the waves are likely dissipated via non-linear effects and/or turbulent damping in regions above the PNS ($r \gtrsim 40$ km). An important timescale is the wave crossing timescale
\beq
\label{tcross}
t_{\rm cross}(r) = \int^r_0 \frac{dr}{v_g}\,\,,
\eeq
where $v_g$ is the wave radial group velocity, $v_g = \partial \omega/\partial k_r$. Outside of the inner core ($r\gtrsim 30$ km), $v_g \approx c_s$ and so the wave crossing time for acoustic waves is approximately
\beq
\label{tcross2}
t_{\rm cross}(r) \approx \int^r_0 \frac{dr}{c_s(r)} \sim t_{\rm dyn} (r)\,\,.
\eeq
The last equality arises because the regions of wave propagation are approximately virialized, so that the sound crossing time is comparable to the dynamical time. Low frequency waves are buoyancy waves in the PNS, and these waves have much larger wave crossing times because the radial group velocity of buoyancy waves is much smaller than that of sound waves. Figure \ref{NStime} shows $t_{\rm cross}$ evaluated at $r=50\,{\rm km}$ as a function of wave frequency. Wave crossing times over this region range are $\sim \! 1\,{\rm ms}$ for high frequency waves.

However, $t_{\rm cross}(r)$ is not always a good estimate for a wave damping timescale because waves can reflect off the PNS surface and become trapped within the PNS. Low frequency waves only gradually tunnel through the overlying evanescent region (cf. Figure \ref{NSwavefunction}), and therefore a more relevant timescale is the wave leakage timescale $t_{\rm leak}(r)$. This timescale reflects the rate at which wave energy leaks out of regions below $r$, and is calculated via Equation \ref{tleak} in Appendix \ref{wavemode}.

Figure \ref{NStime} plots $t_{\rm leak}$, evaluated at $r=50\,{\rm km}$, as a function of wave frequency. For high frequency waves $(f \gtrsim 1\,{\rm kHz})$, $t_{\rm cross} \simeq t_{\rm leak}$ because these waves are not trapped within the PNS. They should therefore damp out on timescales of milliseconds. Lower frequency waves exhibit a series of peaks in the value of $t_{\rm leak}$. These peaks correspond approximately to PNS oscillation modes, for which the waves are mostly reflected at the PNS surface. Waves with frequencies $f \simeq 0.77 \,{\rm kHz}$, corresponding to the PNS f-mode, leak outwards on timescales of $t_{\rm leak} \sim 10\,{\rm ms}$. This timescale is quite similar to the damping times found in simulations (e.g., O12, A14). We therefore conclude that the lifetime of these waves is well approximated by $t_{\rm leak}$. Waves at this frequency are likely damped via turbulent/non-linear effects only after they are able to leak into the outer envelope ($r \gtrsim 30$ km). Waves at lower frequencies corresponding to gravity modes within the PNS may have much longer life times, although we shall see in Section \ref{GW} that their contribution to the GW spectrum is quite small.

\subsection{Rotation}
\label{rotation}

The most obvious impact of rotation is to provide centrifugal support for the PNS and surrounding material. As the angular spin frequency $\Omega$ increases, centrifugal support causes the postbounce structure to be less compact (i.e., lower central densities and larger PNS radii) and more oblate. The change in background structure is well-captured by the simulations A14 used to generate our background structures. We have also attempted to approximately account for the effect of the oblateness in the strength of the forcing that excites the waves (see Section \ref{oscillations}). Because the centrifugally supported PNSs are less compact, their corresponding dynamical frequencies $\Omega_{\rm dyn}$ and mode frequencies $\omega_\alpha$ are generally lower. Including only the effect of rotation on the background structure, we would expect the frequencies of all axisymmetric modes to decrease with increasing rotation.

However, in the wave calculations presented in Appendix \ref{equations}, we have ignored the Coriolis and centrifugal terms in the momentum equations. These terms become very important when the oscillation mode frequency $f$ becomes comparable to $2 f_{\rm spin}$. We are primarily interested in waves with frequencies $f \lesssim 1\,{\rm kHz}$, whereas the inner core spin frequencies of the background models are of order $2 f_{\rm spin} \sim 0.6 \,{\rm kHz}$. Rotation therefore has a strong effect on the wave dynamics and may significantly change our results. Details on the influence of rapid rotation on the oscillation modes of NSs can be found in \cite{bildsten:96}, \cite{dimmelmeier:06}, and \cite{passamonti:09}.

We can attempt to predict the influence of rotation based on perturbation theory. For the axisymmetric oscillation modes of interest, the first order change in mode frequency (proportional to $\Omega$) vanishes. Second order corrections include the Coriolis and centrifugal forces, the non-spherical background structure, and spin-induced coupling between oscillation modes. For axisymmetric $l=2$ f-modes, rotation has only a small effect on the mode frequency \citep{dimmelmeier:06}. For low frequency axisymmetric g-modes, however, rotation typically increases axisymmetric mode frequencies (see \citealt{bildsten:96}, \citealt{lee:97}). The f-mode with $f\simeq 0.8 \,{\rm kHz}$ in Figure \ref{NStime} is somewhat mixed in character. It behaves like an f-mode at $r\lesssim$10 km where it is evanescent, but behaves like a g-mode in the range $10\,{\rm km} \lesssim r \lesssim 20\,{\rm km}$. Its mixed character makes is difficult to easily predict the effect of rotation on its frequency. 

An additional complication is that rotation couples spherical harmonics of $Y_{l,m}$ and $Y_{l\pm2,m}$. Axisymmetric $l\!\!=\!\!2$, $m\!\!=\!\!0$ waves of interest couple to both $l\!\!=\!\!0$ (radial) waves and $l\!\!=\!\!4$ waves. Consequently both $l\!\!=\!\!0$ and $l\!\!=\!\!4$ waves will obtain quadrupole components that allow them to generate GWs. Rotation also induces mode mixing between modes of the same degree $l$, e.g., between the $l\!\!=\!\!2$ PNS f-mode and g-modes.  Rotational mixing between the modes prevents their frequencies from crossing (see Section 4 of \citealt{fuller:14a}), and the modes instead undergo an ``avoided" crossing in which they exchange character. During the avoided crossing, the mode frequency separation is approximately constant, and the modes are superpositions of the unperturbed modes, giving them a hybrid mode character. We revisit the observational consequence of rotational mode coupling in Section \ref{rotation2}.

%From Figure \ref{NSspinGR} we see that the f-mode and g-mode peaks converge in more rapidly rotating models.

Finally, rotation introduces inertial waves/modes, which are restored by the Coriolis force. In uniformly rotating bodies, inertial modes generally exist within the angular frequency range $\omega \lesssim 2 \Omega$. In our most rapidly rotating models, the f-modes and g-modes lie in the range $\omega \lesssim 2 \Omega_c$ (where $\Omega_c$ is the peak angular velocity in the postbounce model). Therefore inertial modes could potentially influence the wave spectrum, either directly (by producing a peak in the GW spectrum) or indirectly (by affecting the frequency of the f-mode).

\subsection{Special and General Relativity}

In this work, we largely ignore the effects of special relativity (SR) and general relativity (GR). Although relevant for the PNS, they greatly complicate the analysis. The effects of GR on NS oscillation modes have been studied extensively (see, e.g., \citealt{thorne:69a,thorne:69b}, \citealt{detweiler:75}, \citealt{cutler:92}, \citealt{andersson:98}, \citealt{lockitch:01,lockitch:03}, \citealt{boutloukos:07}, \citealt{gaertig:09}, \citealt{burgio:11}). These results indicate that we can anticipate GR to affect mode frequencies by at most $\mathcal{O} \big[ 2GM(r)/(r c^2) \big] < 10\%$ at all radii within our model. However, \cite{dimmelmeier:02} compared Newtonian and conformally flat GR CC simulations, finding significant differences in postbounce structure and GW spectra. Since our background structures are generated from GR simulations, we expect that GR effects on wave dynamics will be smaller than the effects of rapid rotation.

SR effects also become important when fluid motions approach the speed of light. The velocity associated with a perturbation with $\xi_r=2\,{\rm km}$ at a frequency $f=1\,{\rm kHz}$ is $\delta v \approx 1.2 \times 10^9\,{\rm cm/s} \approx 4 \times 10^{-2} c$. Therefore the wave amplitudes predicted by our calculations (see Figure \ref{NSnonlin}) generate fluid velocities far below speed of light, and so SR corrections are small.

\section{Gravitational Wave Signatures}
\label{GW}

GW and and neutrino emission are the only way of directly observing the core dynamics of CC SNe \citep{ott:09}. Here we attempt to quantify the GW signatures produced by the bounce-excited oscillations and compare them with simulation results.

\subsection{Gravitational Wave Spectrum}
\label{GWs}

We now turn our attention to GW wave emission induced by the fluid waves. The time-integrated GW energy emitted per unit frequency is 
\begin{align}
\label{GW2}
2 \pi \int dt \ \dot{E}_{{\rm GW},\omega} &= \frac{dE_{\rm GW}}{d f} \nonumber \\
&= \frac{2 G}{5 c^5} \omega^6 |\delta Q_\omega|^2\,\,,
\end{align}
where $\delta Q_\omega$ is the quadrupole moment per unit frequency (calculated from Equation \ref{quad1}) and the second line is from O12. The GW energy corresponds to a characteristic dimensionless wave strain \citep{flanagan:98b}
\beq
\label{strain}
h_{\rm char} = \sqrt{ \frac{2}{\pi^2} \frac{G}{c^3 D^2} \frac{dE_{\rm GW}}{df} }\,\,.
\eeq
We use the fiducial distance $D=10\,{\rm kpc}$ in our presented results.

Caution must be used when evaluating Equation \ref{GW2}. Although the wave frequency is constant in radius, the quadrupole moment $|\delta Q_\omega|$ generally increases at larger radii (see Figure \ref{NSwavefunction}). The GW energy flux is therefore dependent on which radius we choose to evaluate $|\delta Q_\omega|$. Moreover, the total energy emitted by the GWs could be larger than the wave energy, especially for high frequency waves. This unphysicality reflects the fact that we have not taken GW emission into account in the fluid oscillation equations; in reality the waves are attenuated as they emit GWs. 

In what follows, we calculate GW energies and amplitudes with $\delta Q_\omega$ evaluated at $r_{\rm GW}=30\,{\rm km}$. This radius is a good choice as long as the fluid waves damp out at radii just above $r_{\rm GW}$. If they are able to propagate to larger radii, they will obtain larger quadrupole moments (see Figure \ref{NSwavefunction}) and may emit more energy in GWs. Therefore, the energy fluxes we calculate should be viewed as order of magnitude estimates, and only full non-linear hydrodynamic simulations can yield quantitatively reliable predictions. The frequencies of the peaks in the GW spectrum are not strongly affected by our choice of $r_{\rm GW}$ because these peaks are primarily determined by the values of the PNS mode frequencies.

\begin{figure}
\begin{center}
\includegraphics[scale=0.45]{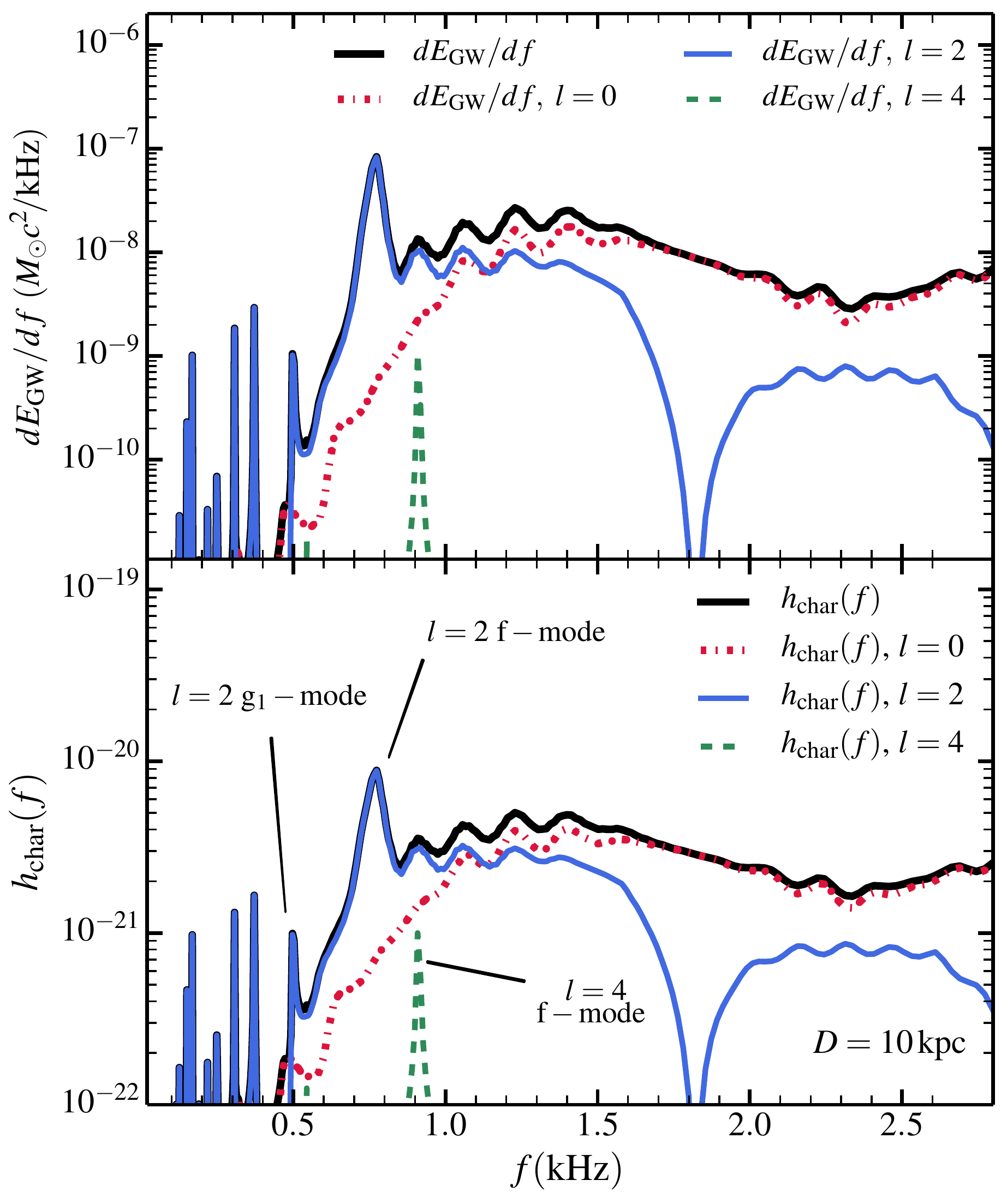}
\end{center} 
\caption{ \label{NSwave} {\bf Top:} GW energy spectrum, $dE_{\rm GW}/df$, due to bounce-excited oscillations, calculated using our A3O04 model (from A14). We have plotted contributions from $l=2$, $l=4$, and $l=0$ waves. The $l=0$ and $l=4$ waves emit GWs because they gain a quadrupole moment due to the aspherical background structure, and their energy spectrum is only an order of magnitude estimate (see text). The broad spectrum of GWs at higher frequencies is created by outgoing pressure waves that create the bounce signal. This signal is likely to be overestimated, as these waves quickly dissipate via shock formation before they can radiate GWs. {\bf Bottom:} Dimensionless characteristic wave strain (Equation \ref{strain}) for the same model, computed at $D=10\,{\rm kpc}$. We have labeled peaks corresponding to some of the PNS oscillation modes.}
\end{figure}

Figure \ref{NSwave} shows a plot of the total energy radiated in GWs per unit frequency and the associated characteristic wave strains. For frequencies $f \lesssim 1$ kHz, the GW energy is sharply peaked around characteristic frequencies. These frequencies essentially correspond to the oscillation mode frequencies of the PNS. The widths of the peaks are the inverse of the mode lifetimes, i.e., the timescale on which the mode energy is able to leak out of the PNS, $t_{\rm leak}$ (see Equation \ref{tleak}). There are no sharp peaks at high frequencies $f \gtrsim 1$ kHz because these frequencies correspond to pressure waves (which are not reflected at the edge of the PNS) that quickly propagate outwards on a wave crossing time $t_{\rm cross}$. 

The peak at $f\simeq 0.8$ kHz corresponds to the $l\!\!=\!\!2$ axisymmetric PNS oscillation mode (see Section \ref{waveprop}). This mode contains more energy than any other because it couples well (in both physical and frequency space) to the forcing produced at bounce. We therefore confirm the hypothesis of O12 that the peak centered at $f \sim 0.8$ kHz in their GW spectra is generated by the axisymmetric quadrupolar PNS f-mode. The f-mode is expected to dominate the early postbounce GW signature of rapidly rotating CC SNe. The peaks at lower frequencies correspond to $l\!\!=\!\!2$ PNS g-modes; the first is the $g_1$-mode at $f\simeq 0.5$ kHz. We reiterate that rotational effects are important and may substantially alter mode frequencies, especially in the low frequency part of the spectrum. 

We have also attempted to calculate the GW spectra due to quasi-radial ($l\!\!=\!\!0$) and $l\!\!=\!\!4$ waves, which obtain quadrupole moments due to rotational mixing with $l\!\!=\!\!2$ modes (see Section \ref{rotation}) and due to the centrifugally distorted background structure. To estimate GW spectra for $l\!\!=\!\!0$ waves, we remove the $\epsilon=(\Omega/\Omega_{\rm dyn})^2$ dependence of the forcing term (see Equation \ref{fval}) because centrifugal distortion is not needed to excite radial waves. To calculate their quadrupole moment, we multiply the right hand side of Equation \ref{quad2} by $\epsilon$, which approximately accounts for the quadrupole moment of the background structure and allows the $l\!\!=\!\!0$ waves to emit GW. For $l\!\!=\!\!4$ waves, we replace $\epsilon$ with $\epsilon^2$ in Equation \ref{fval} to estimate the reduced strength of the wave excitation, and use the same procedure as $l=0$ waves to calculate an approximate quadrupole moment. This procedure is rudimentary and should not be expected to yield accurate quantitative predictions for the energy radiated by $l=0$ and $l=4$ waves, although it can be used for a qualitative understanding.

The GW spectrum produced by the $l\!\!=\!\!0$ waves is a smooth continuum rather than being peaked at mode frequencies. The reason is that $l\!\!=\!\!0$ waves are not well reflected from the PNS edge, and so energy in $l\!\!=\!\!0$ waves leaks out of the PNS on a wave crossing (dynamical) timescale. Moreover, $l\!\!=\!\!0$ g-modes do not exist, instead low frequency $l\!\!=\!\!0$ waves are evanescent in the PNS when $\omega<N$. This is in stark contrast to the $l\!\!=\!\!2$ waves, which can be trapped in regions with $\omega<N$ to form oscillation modes. Instead, the force exherted by the bounce is transferred to $l\!\!=\!\!0$ waves of a broad range in frequencies, which quickly travel outward and steepen into shocks; it is this process which generates the outgoing shock created by bounce.  Although the $l\!\!=\!\!0$ waves are important for the GW spectrum of Figure \ref{NSwave} for $f\gtrsim 1\,{\rm kHz}$, their GW strain peaks near bounce and contributes primarily to the bounce signal (see Figure \ref{NSstrain} and Section \ref{intro}). The same is true for higher frequency ($f\gtrsim 1\,{\rm kHz}$) quadrupolar waves. After bounce, the high frequency waves are quickly dissipated by shocks, and the $l\!\!=\!\!2$ modes dominate the GW spectrum. This idea is consistent with the results of K15, who find the GW spectrum is more strongly peaked around mode frequencies when the bounce is windowed out.

The $l\!\!=\!\!4$ waves are less efficiently excited by the bounce than $l\!\!=\!\!2$ waves, but the response is similarly peaked around mode frequencies. The largest peak is centered around the $l\!\!=\!\!4$ f-mode at $f \approx 1 \, {\rm kHz}$, although we expect this mode to radiate considerably less GW energy than the $l\!\!=\!\!2$ f-mode. Nonetheless, given our rudimentary methods, we speculate that the $l\!\!=\!\!4$ f-mode may be detectable, especially for very rapidly rotating progenitors.

\subsection{Comparison with Non-linear Simulations}
\label{observation}

We now compare our semi-analytical results with the simulation results of O12 and A14. The GW energy spectra of these simulations generally contain a few distinguishing features: 
\\ {\it 1.} A prominent peak of maximum GW energy in the range $0.7 \,{\rm kHz} \lesssim f_{\rm max} \lesssim 0.8 \,{\rm kHz}$. \\ {\it 2.} A broad spectrum of GW energy at frequencies $f \lesssim 2\,{\rm kHz}$. 
%\\ {\it 3.} In K15, there is a less prominent peak at $1.5 \,{\rm kHz} \lesssim f_2 \lesssim 1.6 \,{\rm kHz}$, and small peaks around $f\sim 0.5\,{\rm kHz}$ and $f\sim 0.95 \,{\rm kHz}$.

The main peak at $f_{\rm max} \sim 0.75$ kHz is due to the $l\!\!=\!\!2$, $m\!\!=\!\!0$ f-mode of the PNS, as speculated by O12. However, both the frequency and wave function of this mode are likely influenced by rotational interaction with other modes (see Section \ref{rotation} and discussion below). The g$_1$-mode may be responsible for peaks near $f\sim 0.5$ kHz, and the $l\!\!=\!\!4$ f-mode may produce a peak at $f\sim 0.95$ kHz. Our linear calculations do not easily account for any peaks at $f \gtrsim 1$ kHz. We speculate that peaks near $f \sim 1.5\,{\rm kHz}$ are due to the first harmonic of $f_{\rm max}$, generated due to the non-linearity of the f-mode responsible for $f_{\rm max}$ (see Appendix \ref{nonlinear}). 

The broad spectrum of GW energy (visible as broad peaks near $f\sim1.2\,{\rm kHz}$ in Figure \ref{NSwave} and the top panel of Figure \ref{NSspinGR}) is produced by both quasi-radial and quadrupolar waves which are not efficiently reflected and quickly propagate out of the PNS. However, we caution that our methods may overpredict the GW signal from these waves, as they quickly steepen into shocks before generating GWs. This may account for the lack of a very broad peak at $f\sim1.2\,{\rm kHz}$ in the rapidly rotating simulations of O12 and A14 (lower panel of Figure \ref{NSspinGR}).

Our calculations predict total GW energy outputs and wave strains roughly consistent with the results of O12 and A14 when we use a forcing strength of $A\sim 2$ (Equation \ref{fval}), although there are significant uncertainties in calculating the GW spectrum. Here, we claim only that our method produces a sensible order of magnitude estimate for GW energies, and that it provides a physical explanation for some of the features in the GW spectra from rotating core collapse simulations. 

Finally, we comment on the widths of the GW spectral peaks. As discussed in Section \ref{nonad}, the damping timescale for the oscillation modes is determined by the wave leakage timescale into the envelope. For the $l\!\!=\!\!2$ f-mode, this leakage time is $\sim 10$ ms (in good agreement with the GW decay timescale seen in O12 and A14), corresponding to the width of $\Delta f \sim 0.1\,$kHz for the peak at $f_{\rm max}$. The g-mode peaks in Figure \ref{NSwave} are narrower on account of the long leakage timescale for the g-modes, but their widths are underestimated since the g-modes may be damped via non-linear processes or modified by the background structural evolution. 

%The remnant evolution rate of $\sim 30$ ms also sets a lower limit of $\sim 30$ Hz for the width of any feature in the spectrum, therefore the spectral peaks at low frequencies ($f \lesssim 300$ Hz) will be largely smoothed out. 

\subsubsection{Rotation}
\label{rotation2}

\begin{figure}
\begin{center}
\includegraphics[scale=0.39]{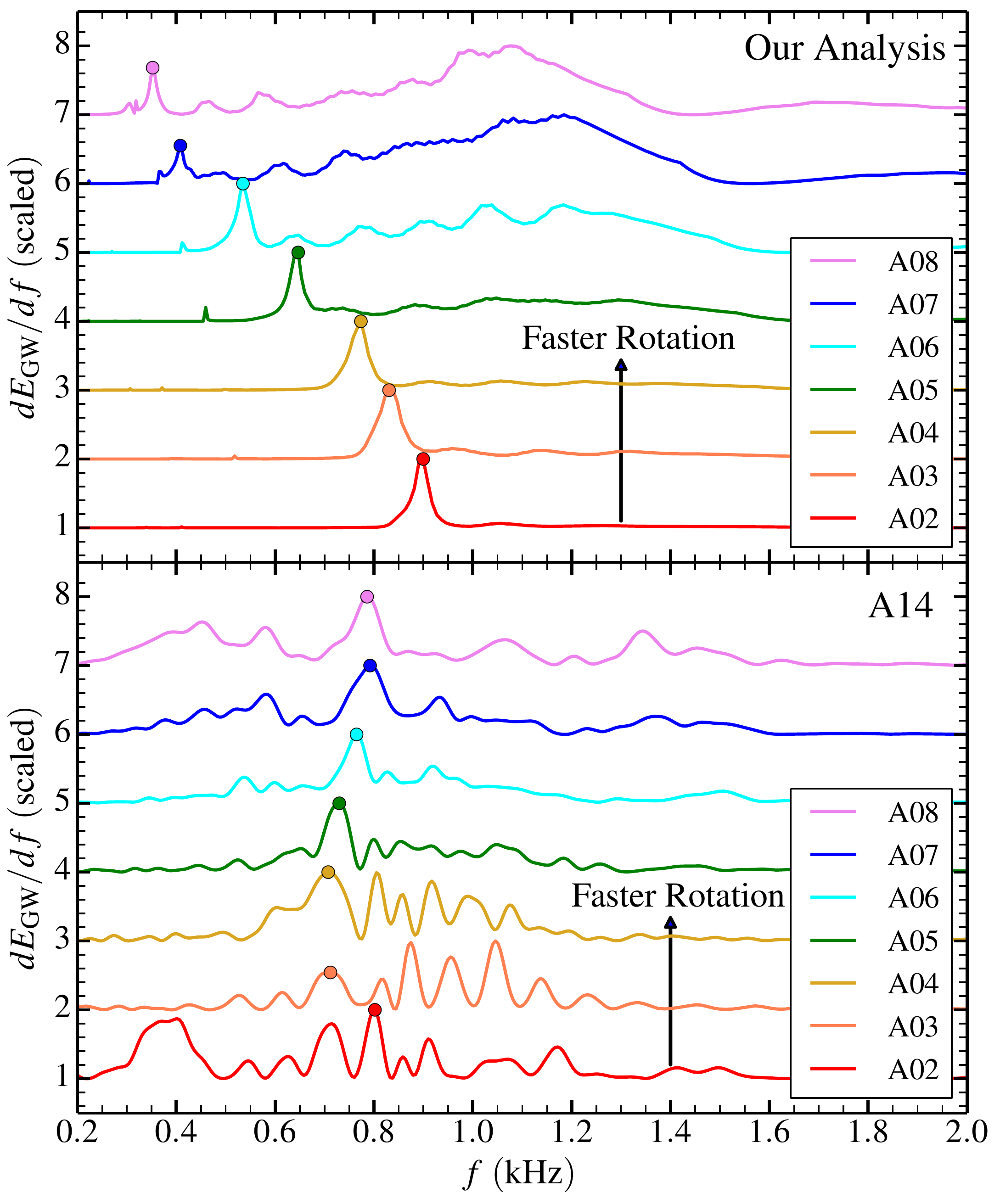}
\end{center} 
\caption{ \label{NSspinGR} {\bf Top:} Our computed GW spectra for the sequence of models A3O02-A3O08 of A14. Larger model numbers are more rapidly rotating. The spectrum of each model has been scaled to its maximum amplitude and shifted vertically for clarity. The f-mode corresponds to the sharp peak in each curve, and is marked with a circle symbol. The broad humps of power at higher frequencies are due to outgoing pressure waves which create the bounce signal, although their contribution to the GW spectrum is likely overestimated (see text). {\bf Bottom:} GW energy spectra from the same simulations (A14) used to generate the background models for our calculations. The peak which we speculate to be generated by the quadrupolar f-mode has been marked with a circle symbol, although its location is ambiguous in slowly rotating models. The mismatch in frequency of the f-mode peak between top and bottom panels is most likely due to the strong effect of rotation on the wave dynamics (see text), which is not included our calculations. }
\end{figure}

A14 computed GW energy spectra of rapidly rotating progenitors as a function of the rotation rate. To compare with their results, we generate background structures and predicted GW spectra from the simulations of A14 for different rotation rates, using the same techniques described in the preceding sections.

Figure \ref{NSspinGR} displays our computed GW spectra (top panel) for the rapidly rotating models of A14, in addition to the GW spectra obtained from the simulations themselves (bottom panel). As the rotation rate increases, the PNS and surrounding material have more centrifugal support and become less compact, decreasing their dynamical frequency, Lamb frequencies $L_2$, and Brunt-V\"{a}is\"{a}l\"{a} frequencies $N$. For this reason, the prominent peaks computed from our semi-analytical analysis shift to lower frequencies with increasing rotation rate. They also decrease in power due to the strong dependence of $\dot{E}_{\rm GW}$ on frequency (see Equation \ref{GW2}). Rapidly rotating models also contain significant power in a broad peak centered near $f\sim\,1.2 \, {\rm kHz}$. This power is created by pressure waves which quickly leak into the envelope, although this power may be overestimated (see discussion above).

A monotonic trend of decreasing $f_{\rm max}$ with spin frequency is {\it not} observed in A14. Instead, their results indicate a more complex dependence in which $f_{\rm max}$ increases in frequency at moderate rotation rates and decreases at very high rotation rates, although over most of the parameter space $0.7 \, {\rm kHz} < f_{\rm max} < 0.8 \, {\rm kHz}$.

We attribute the difference to our neglect of the effects of the Coriolis and centrifugal forces on the wave dynamics, although GR effects could have some influence as well. As discussed in Section \ref{rotation}, the f-mode is really a hybrid f-mode/g-mode that propagates near the sharp density gradient at the interface between the PNS and surrounding material. Due to its g-mode component, the Coriolis force will tend to increase its frequency \citep{bildsten:96} in rapidly rotating models. We speculate that the increasing impact of the Coriolis force with rotation largely cancels the effect of the less compact structure, causing $0.7 \, {\rm kHz} < f_{{\rm f-mode}} < 0.8 \, {\rm kHz}$ over a large range of spin frequencies. A fairly weak dependence on spin frequency is also seen for axisymmetric $l=2$ f-modes in \cite{dimmelmeier:06}.

Rotational mixing between the $l\!\!=\!\!2$ f-mode and g$_1$-mode may also affect the frequency behavior of the GW spectra. This mixing can generate mode ``repulsion" between the f-mode and g-mode during an avoided crossing, which may increase the value of $f_{{\rm f-mode}}$ at some spin rates. Moreover, in very rapidly rotating models for which $\omega_{{\rm f-mode}} \lesssim 2 \Omega_c$, mixing with inertial modes may also influence the wave spectrum. 

We also note that O12 find an oscillation in the central density $\rho_c$ of their simulations with frequency $f_{\rm max}$. If the oscillation at $f_{\rm max}$ is generated by a purely quadrupolar mode, we should expect a negligible oscillation in $\rho_c$ because all perturbations vanish at the center for modes with $l\geq2$. However, as described above, rotation induces mixing between waves of $l$ and $l\pm2$. In rapidly rotating models the quadrupole waves thus obtain a radial component which will generate oscillations in $\rho_c$. 

Properly incorporating the influence of rotation is beyond the scope of this paper. We hope that future studies with more rigorous implementations of rotational effects will improve our understanding of their influence on the GW spectrum. Unfortunately, the rather complex dependence of $f_{\rm max}$ on rotation frequency may complicate interpretations of a GW signal from a nearby CC SNe. Simulation results (which naturally include effects of rotation, non-linearity, etc.) such as those of A14 (and the upcoming study by K15) may therefore be vital for interpretation of observational results.

\subsection{Neutrino Signatures}

Neutrinos offer another opportunity to observe the interior dynamics of CC SNe. Our calculations do not include any effects of neutrinos, but we may speculate on the neutrino signature of bounce-excited oscillations. The oscillations generate perturbations in density in temperature which may generate oscillations in neutrino generation rates. The oscillations may also perturb the effective location of the neutrinosphere. Both these effects may generate oscillations in the observed neutrino flux, with identical frequencies to the peaks in the GR spectrum. Indeed, O12 find oscillations in the neutrino flux that are correlated with the GR signal. We do not investigate the physics of these oscillations in detail, but merely speculate that we expect a peak in the Fourier transform of a neutrino light curve at $0.7  \,{\rm kHz} \lesssim f \lesssim 0.8 \,{\rm kHz}$.

\section{Discussion and Conclusions}
\label{disc}

We have examined the physics of wave generation, propagation, and dissipation in rapidly rotating core-collapse supernovae (CC SNe). Our linear, semi-analytic methods complement recent hydrodynamic simulations of rotating CC SNe (O12 and A14), and they shed light on the basic physics at play. The quadrupolar distortion of the centrifugally distorted PNS inner core generates axisymmetric quadrupolar ($l=2$) waves at bounce. Because of their quadrupolar nature, these waves can generate a strong gravitational wave (GW) signal that could be detected by advanced LIGO for a galactic CC SN (O12, A14).

The GW signal is largely determined by two factors: the strength of the wave excitation provided by core bounce, and the ensuing propagation/dissipation of the waves. The strength of the force (per unit mass) provided by the bounce is roughly proportional to the local gravitational acceleration (see Equation \ref{f1}), therefore, the wave excitation is spatially confined to the inner core of the PNS ($r \lesssim 30\,{\rm km}$). The duration of the force is approximately the dynamical time of the inner core of the PNS, $t_{\rm dyn} \sim 1\,{\rm ms}$. Hence, wave excitation is peaked around frequencies $f \sim t_{\rm dyn}^{-1} \sim 1\,{\rm kHz}$. For quadrupolar waves, the strength of the forcing is also proportional to the centrifugal distortion, $\epsilon$, of the inner core of the PNS, and therefore more rapidly rotating progenitors typically transfer more power into quadrupolar oscillations.

The behavior of quasi-radial ($l\!=\!0$) and quadrupolar ($l\!=\!2$) bounce excited waves is fundamentally different. Unlike quadrupolar waves, the quasi-radial waves are not reflected at the edge of the PNS and quickly propagate into the overlying envelope where they generate the bounce-induced shock. Although the GW bounce signal (see Figure \ref{NSstrain}) contains a contribution from the quasi-radial bounce of the PNS (which has a quadrupolar component because of the centrifugal distortion of the PNS), quasi-radial waves contribute little to the GW ring down signal after bounce.

In contrast, quadrupolar waves propagate as buoyancy waves within the PNS in regions where ($\omega<N$). These waves are efficiently reflected at the edge of the PNS ($r\sim 30\,{\rm km}$), and become trapped within the PNS. In-going and outgoing waves interfere to create standing waves, or oscillation ``modes", of the PNS. Consequently, the quadrupolar wave energy is peaked around characteristic ``mode" frequencies. The wave energy leaks out of the PNS on timescales of $\sim 10\,{\rm ms}$, propagating outwards into the envelope as acoustic waves, and eventually dissipating.

The largest early postbounce ($t-t_{\rm bounce} \lesssim 20 \, {\rm ms}$) GW signal is produced by the fundamental quadrupolar oscillation ``mode" of the PNS, whose frequency is typically $f\sim 0.75 \,{\rm kHz}$. This f-mode is essentially a surface wave of the PNS, and owes its existence to the sharp density gradient at the edge of the PNS. However, we emphasize that the f-mode has some buoyancy wave (g-mode) character because it propagates in a region where $\omega<N$. Moreover, the f-mode energy leaks into the low density surrounding envelope, and its properties are not accurately captured by a calculation of oscillation modes in an isolated PNS or NS. Weaker postbounce GW signals may be produced by higher order $l\!=\!2$ g-modes or the $l\!=\!4$ f-mode (like the quasi-radial $l\!=\!0$ waves, the $l\!=\!4$ waves obtain a quadrupole moment due to the centrifugal distortion of the background structure). Finally, the first harmonic of the f-mode, at a frequency $f\sim 1.5\,{\rm kHz}$, may appear in GW spectra due to the somewhat non-linear amplitude of the f-mode. 

The greatest uncertainty and source of error in our analysis is the neglect of the effects of Coriolis and centrifugal forces on the wave dynamics. Indeed, these forces are quite important since typical angular spin frequencies are comparable to the wave frequencies. Our calculations (incorrectly) predict that the quadrupolar f-mode frequency monotonically decreases with increasing progenitor spin frequency due to the less compact background structure. In contrast, simulations (A14, K15) show only a weak dependence of the f-mode frequency on the spin frequency. This suggests that rotation helps increase the f-mode frequency and largely offsets the frequency decrease due to lower average density in more rapidly rotating models. We speculate that the Coriolis force increases the f-mode frequency, as it does to g-modes in rapidly rotating stars (\citealt{bildsten:96}, \citealt{lee:97}). In any case, the bounce signal likely provides a better indication of progenitor spin frequency than the f-mode frequency (A14). 

Nonetheless, the detection of both the bounce signal and the f-mode would provide useful constraints on the structure and spin frequency of the inner core. The frequency of the f-mode varies little with rotation rate and depends primarily upon the equation of state (A14, K15), while the bounce signal is sensitive to both equation of state and spin frequency. Measuring $f_{\rm f-mode}$ could thus break a partial degeneracy between the compactness of the inner core and its spin frequency. Additionally, for slow-moderate rotation rates, the amplitude $|a_{\rm f-mode}|$ of the quadrupolar f-mode signal is roughly proportional to the square of the core spin frequency, $\Omega^2$. A measurement of $|a_{\rm f-mode}|$, coupled with an accurate distance measurement (likely to be fairly well-determined for a galactic CC SN), will thus provide an additional constraint on inner core spin frequency.

The simple analysis presented here has shed some light on wave dynamics within rapidly rotating CC SNe, and how the basic properties of the progenitor contribute to the GW spectrum produced by waves excited at bounce. However, our analysis is not sufficient to accurately calculate GW spectra suitable for comparison with observations. We encourage further simulations of rapidly rotating CC SNe that generate GW spectra over reasonable ranges in parameter space (e.g., using a range of rotation frequencies, differential rotation profiles, equations of state, neutrino approximations, etc.) complementary to those already performed (e.g., \citealt{dimmelmeier:08,abdikamalov:10}, O12, A14). These simulations will provide templates for comparison with the GWs observed by advanced LIGO in the event of a galactic CC SN, and they will therefore be an important tool for understanding the extreme physics of these explosions.

\section*{Acknowledgments} 

We thank Nick Stergioulas for helpful comments. This work was partially supported by the National Science Foundation
under award nos. AST-1205732, PHY-1125915, and PHY-1151197, by a Lee
DuBridge Fellowship awarded to JF at Caltech, and by the Sherman
Fairchild Foundation. Some of the non-linear hydrodynamics simulations
for this study were carried out on the Caltech compute cluster Zwicky,
which is funded by NSF MRI-R2 award no.\ PHY-0960291, and on NSF XSEDE
resources under allocation TG-PHY100033.

\bibliography{bibliography/jet_references,bibliography/bh_formation_references,bibliography/gw_references,bibliography/sn_theory_references,bibliography/grb_references,bibliography/nu_obs_references,bibliography/methods_references,bibliography/eos_references,bibliography/NSNS_NSBH_references,bibliography/stellarevolution_references,bibliography/nucleosynthesis_references,bibliography/gr_references,bibliography/sn_observation_references,bibliography/numrel_references,bibliography/gw_data_analysis_references,bibliography/ns_references,bibliography/stellar_oscillations_references,bibliography/gw_detector_references}

\begin{thebibliography}{}
\makeatletter
\relax
\def\mn@urlcharsother{\let\do\@makeother \do\$\do\&\do\#\do\^\do\_\do\%\do\~}
\def\mn@doi{\begingroup\mn@urlcharsother \@ifnextchar [ {\mn@doi@}
  {\mn@doi@[]}}
\def\mn@doi@[#1]#2{\def\@tempa{#1}\ifx\@tempa\@empty \href
  {http://dx.doi.org/#2} {doi:#2}\else \href {http://dx.doi.org/#2} {#1}\fi
  \endgroup}
\def\mn@eprint#1#2{\mn@eprint@#1:#2::\@nil}
\def\mn@eprint@arXiv#1{\href {http://arxiv.org/abs/#1} {{\tt arXiv:#1}}}
\def\mn@eprint@dblp#1{\href {http://dblp.uni-trier.de/rec/bibtex/#1.xml}
  {dblp:#1}}
\def\mn@eprint@#1:#2:#3:#4\@nil{\def\@tempa {#1}\def\@tempb {#2}\def\@tempc
  {#3}\ifx \@tempc \@empty \let \@tempc \@tempb \let \@tempb \@tempa \fi \ifx
  \@tempb \@empty \def\@tempb {arXiv}\fi \@ifundefined
  {mn@eprint@\@tempb}{\@tempb:\@tempc}{\expandafter \expandafter \csname
  mn@eprint@\@tempb\endcsname \expandafter{\@tempc}}}

\bibitem[\protect\citeauthoryear{{Aasi et al.\ (LIGO Scientific
  Collaboration)}}{{Aasi et al.\ (LIGO Scientific
  Collaboration)}}{2014}]{aligo}
{Aasi et al.\ (LIGO Scientific Collaboration)} J.,  2014, submitted to \cqg \
  arXiv:1411.4547, \href {http://adsabs.harvard.edu/abs/2014arXiv1411.4547T} {}

\bibitem[\protect\citeauthoryear{{Abdikamalov}, {Ott}, {Rezzolla}, {Dessart},
  {Dimmelmeier}, {Marek}  \& {Janka}}{{Abdikamalov}
  et~al.}{2010}]{abdikamalov:10}
{Abdikamalov} E.~B.,  {Ott} C.~D.,  {Rezzolla} L.,  {Dessart} L.,
  {Dimmelmeier} H.,  {Marek} A.,   {Janka} H.,  2010, \prd, \href
  {http://adsabs.harvard.edu/abs/2010PhRvD..81d4012A} {81, 044012}

\bibitem[\protect\citeauthoryear{{Abdikamalov}, {Gossan}, {DeMaio}  \&
  {Ott}}{{Abdikamalov} et~al.}{2014}]{abdikamalov:14}
{Abdikamalov} E.,  {Gossan} S.,  {DeMaio} A.~M.,   {Ott} C.~D.,  2014, \mn@doi
  [\prd] {10.1103/PhysRevD.90.044001}, \href
  {http://adsabs.harvard.edu/abs/2014PhRvD..90d4001A} {90, 044001}

\bibitem[\protect\citeauthoryear{{Andersson}}{{Andersson}}{1998}]{andersson:98}
{Andersson} N.,  1998, \apj, 502, 708

\bibitem[\protect\citeauthoryear{{Bildsten}, {Ushomirsky}  \&
  {Cutler}}{{Bildsten} et~al.}{1996}]{bildsten:96}
{Bildsten} L.,  {Ushomirsky} G.,   {Cutler} C.,  1996, \mn@doi [\apj]
  {10.1086/177012}, \href {http://adsabs.harvard.edu/abs/1996ApJ...460..827B}
  {460, 827}

\bibitem[\protect\citeauthoryear{{Boutloukos} \& {Nollert}}{{Boutloukos} \&
  {Nollert}}{2007}]{boutloukos:07}
{Boutloukos} S.,  {Nollert} H.-P.,  2007, \mn@doi [\prd]
  {10.1103/PhysRevD.75.043007}, \href
  {http://adsabs.harvard.edu/abs/2007PhRvD..75d3007B} {75, 043007}

\bibitem[\protect\citeauthoryear{{Burgio}, {Ferrari}, {Gualtieri}  \&
  {Schulze}}{{Burgio} et~al.}{2011}]{burgio:11}
{Burgio} G.~F.,  {Ferrari} V.,  {Gualtieri} L.,   {Schulze} H.-J.,  2011,
  \mn@doi [\prd] {10.1103/PhysRevD.84.044017}, \href
  {http://adsabs.harvard.edu/abs/2011PhRvD..84d4017B} {84, 044017}

\bibitem[\protect\citeauthoryear{{Burrows}, {Dessart}, {Livne}, {Ott}  \&
  {Murphy}}{{Burrows} et~al.}{2007}]{burrows:07b}
{Burrows} A.,  {Dessart} L.,  {Livne} E.,  {Ott} C.~D.,   {Murphy} J.,  2007,
  \apj, 664, 416

\bibitem[\protect\citeauthoryear{{Cutler} \& {Lindblom}}{{Cutler} \&
  {Lindblom}}{1992}]{cutler:92}
{Cutler} C.,  {Lindblom} L.,  1992, \mn@doi [\apj] {10.1086/170968}, \href
  {http://adsabs.harvard.edu/abs/1992ApJ...385..630C} {385, 630}

\bibitem[\protect\citeauthoryear{{Detweiler}}{{Detweiler}}{1975}]{detweiler:75}
{Detweiler} S.~L.,  1975, \mn@doi [\apj] {10.1086/153504}, \href
  {http://adsabs.harvard.edu/abs/1975ApJ...197..203D} {197, 203}

\bibitem[\protect\citeauthoryear{{Dimmelmeier}, {Font}  \&
  {M{\"u}ller}}{{Dimmelmeier} et~al.}{2002}]{dimmelmeier:02}
{Dimmelmeier} H.,  {Font} J.~A.,   {M{\"u}ller} E.,  2002, \aap, 393, 523

\bibitem[\protect\citeauthoryear{{Dimmelmeier}, {Novak}, {Font},
  {Ib{\'a}{\~n}ez}  \& {M{\"u}ller}}{{Dimmelmeier}
  et~al.}{2005}]{dimmelmeier:05}
{Dimmelmeier} H.,  {Novak} J.,  {Font} J.~A.,  {Ib{\'a}{\~n}ez} J.~M.,
  {M{\"u}ller} E.,  2005, \prd, 71, 064023

\bibitem[\protect\citeauthoryear{{Dimmelmeier}, {Stergioulas}  \&
  {Font}}{{Dimmelmeier} et~al.}{2006}]{dimmelmeier:06}
{Dimmelmeier} H.,  {Stergioulas} N.,   {Font} J.~A.,  2006, \mnras, 368, 1609

\bibitem[\protect\citeauthoryear{{Dimmelmeier}, {Ott}, {Janka}, {Marek}  \&
  {M{\"u}ller}}{{Dimmelmeier} et~al.}{2007}]{dimmelmeier:07}
{Dimmelmeier} H.,  {Ott} C.~D.,  {Janka} H.-T.,  {Marek} A.,   {M{\"u}ller} E.,
   2007, \prl, 98, 251101

\bibitem[\protect\citeauthoryear{{Dimmelmeier}, {Ott}, {Marek}  \&
  {Janka}}{{Dimmelmeier} et~al.}{2008}]{dimmelmeier:08}
{Dimmelmeier} H.,  {Ott} C.~D.,  {Marek} A.,   {Janka} H.-T.,  2008, \prd, 78,
  064056

\bibitem[\protect\citeauthoryear{{Ferrari}, {Miniutti}  \& {Pons}}{{Ferrari}
  et~al.}{2003}]{ferrari:03}
{Ferrari} V.,  {Miniutti} G.,   {Pons} J.~A.,  2003, \mnras, 342, 629

\bibitem[\protect\citeauthoryear{{Ferrari}, {Gualtieri}, {Pons}  \&
  {Stavridis}}{{Ferrari} et~al.}{2004}]{ferrari:04}
{Ferrari} V.,  {Gualtieri} L.,  {Pons} J.~A.,   {Stavridis} A.,  2004, \mnras,
  350, 763

\bibitem[\protect\citeauthoryear{{Flanagan} \& {Hughes}}{{Flanagan} \&
  {Hughes}}{1998}]{flanagan:98b}
{Flanagan} {\'E}.~{\'E}.,  {Hughes} S.~A.,  1998, \mn@doi [\prd]
  {10.1103/PhysRevD.57.4566}, \href
  {http://adsabs.harvard.edu/abs/1998PhRvD..57.4566F} {57, 4566}

\bibitem[\protect\citeauthoryear{{Fuller}, {Lai}  \& {Storch}}{{Fuller}
  et~al.}{2014}]{fuller:14a}
{Fuller} J.,  {Lai} D.,   {Storch} N.~I.,  2014, \mn@doi [\icarus]
  {10.1016/j.icarus.2013.11.022}, \href
  {http://adsabs.harvard.edu/abs/2014Icar..231...34F} {231, 34}

\bibitem[\protect\citeauthoryear{{Gaertig} \& {Kokkotas}}{{Gaertig} \&
  {Kokkotas}}{2009}]{gaertig:09}
{Gaertig} E.,  {Kokkotas} K.~D.,  2009, \mn@doi [\prd]
  {10.1103/PhysRevD.80.064026}, \href
  {http://adsabs.harvard.edu/abs/2009PhRvD..80f4026G} {80, 064026}

\bibitem[\protect\citeauthoryear{{Heger}, {Woosley}  \& {Spruit}}{{Heger}
  et~al.}{2005}]{heger:05}
{Heger} A.,  {Woosley} S.~E.,   {Spruit} H.~C.,  2005, \apj, 626, 350

\bibitem[\protect\citeauthoryear{{Janka}}{{Janka}}{2001}]{janka:01}
{Janka} H.-T.,  2001, \mn@doi [\aap] {10.1051/0004-6361:20010012}, 368, 527

\bibitem[\protect\citeauthoryear{{Kotake}}{{Kotake}}{2013}]{kotake:13review}
{Kotake} K.,  2013, \mn@doi [Comptes Rendus Physique]
  {10.1016/j.crhy.2013.01.008}, \href
  {http://adsabs.harvard.edu/abs/2013CRPhy..14..318K} {14, 318}

\bibitem[\protect\citeauthoryear{{Kotake}, {Yamada}  \& {Sato}}{{Kotake}
  et~al.}{2003}]{kotake:03}
{Kotake} K.,  {Yamada} S.,   {Sato} K.,  2003, \prd, 68, 044023

\bibitem[\protect\citeauthoryear{{Kuroda}, {Takiwaki}  \& {Kotake}}{{Kuroda}
  et~al.}{2014}]{kuroda:14}
{Kuroda} T.,  {Takiwaki} T.,   {Kotake} K.,  2014, \mn@doi [\prd]
  {10.1103/PhysRevD.89.044011}, \href
  {http://adsabs.harvard.edu/abs/2014PhRvD..89d4011K} {89, 044011}

\bibitem[\protect\citeauthoryear{{Langer}}{{Langer}}{2012}]{langer:12}
{Langer} N.,  2012, \araa, 50, 107

\bibitem[\protect\citeauthoryear{{Lee} \& {Saio}}{{Lee} \&
  {Saio}}{1997}]{lee:97}
{Lee} U.,  {Saio} H.,  1997, \apj, \href
  {http://adsabs.harvard.edu/abs/1997ApJ...491..839L} {491, 839}

\bibitem[\protect\citeauthoryear{{Lockitch}, {Andersson}  \&
  {Friedman}}{{Lockitch} et~al.}{2001}]{lockitch:01}
{Lockitch} K.~H.,  {Andersson} N.,   {Friedman} J.~L.,  2001, \mn@doi [\prd]
  {10.1103/PhysRevD.63.024019}, \href
  {http://adsabs.harvard.edu/abs/2001PhRvD..63b4019L} {63, 024019}

\bibitem[\protect\citeauthoryear{{Lockitch}, {Friedman}  \&
  {Andersson}}{{Lockitch} et~al.}{2003}]{lockitch:03}
{Lockitch} K.~H.,  {Friedman} J.~L.,   {Andersson} N.,  2003, \mn@doi [\prd]
  {10.1103/PhysRevD.68.124010}, \href
  {http://adsabs.harvard.edu/abs/2003PhRvD..68l4010L} {68, 124010}

\bibitem[\protect\citeauthoryear{{M{\" u}ller}, {Rampp}, {Buras}, {Janka}  \&
  {Shoemaker}}{{M{\" u}ller} et~al.}{2004}]{mueller:04}
{M{\" u}ller} E.,  {Rampp} M.,  {Buras} R.,  {Janka} H.-T.,   {Shoemaker}
  D.~H.,  2004, \apj, 603, 221

\bibitem[\protect\citeauthoryear{M{\"o}nchmeyer, Sch{\"a}fer, M{\"u}ller  \&
  Kates}{M{\"o}nchmeyer et~al.}{1991}]{moenchmeyer:91}
M{\"o}nchmeyer R.,  Sch{\"a}fer G.,  M{\"u}ller E.,   Kates R.,  1991, \aap,
  246, 417

\bibitem[\protect\citeauthoryear{{M{\"o}sta} et~al.,}{{M{\"o}sta}
  et~al.}{2014}]{moesta:14b}
{M{\"o}sta} P.,  et~al., 2014, \apjl, \href
  {http://adsabs.harvard.edu/abs/2014ApJ...785L..29M} {785, L29}

\bibitem[\protect\citeauthoryear{{M\"uller}}{{M\"uller}}{1982}]{mueller:82}
{M\"uller} E.,  1982, \aap, 114, 53

\bibitem[\protect\citeauthoryear{{M\"uller} \& {Janka}}{{M\"uller} \&
  {Janka}}{1997}]{mueller:97}
{M\"uller} E.,  {Janka} H.-T.,  1997, \aap, 317, 140

\bibitem[\protect\citeauthoryear{{M{\"u}ller}, {Janka}  \&
  {Wongwathanarat}}{{M{\"u}ller} et~al.}{2012}]{mueller:e12}
{M{\"u}ller} E.,  {Janka} H.-T.,   {Wongwathanarat} A.,  2012, \mn@doi [\aap]
  {10.1051/0004-6361/201117611}, \href
  {http://adsabs.harvard.edu/abs/2012A%26A...537A..63M} {537, A63}

\bibitem[\protect\citeauthoryear{{M{\"u}ller}, {Janka}  \&
  {Marek}}{{M{\"u}ller} et~al.}{2013}]{mueller:13gw}
{M{\"u}ller} B.,  {Janka} H.-T.,   {Marek} A.,  2013, \mn@doi [\apj]
  {10.1088/0004-637X/766/1/43}, \href
  {http://adsabs.harvard.edu/abs/2013ApJ...766...43M} {766, 43}

\bibitem[\protect\citeauthoryear{{Murphy}, {Ott}  \& {Burrows}}{{Murphy}
  et~al.}{2009}]{murphy:09}
{Murphy} J.~W.,  {Ott} C.~D.,   {Burrows} A.,  2009, \apj, \href
  {http://adsabs.harvard.edu/abs/2009ApJ...707.1173M} {707, 1173}

\bibitem[\protect\citeauthoryear{{Obergaulinger}, {Aloy}, {Dimmelmeier}  \&
  {M{\"u}ller}}{{Obergaulinger} et~al.}{2006}]{obergaulinger:06b}
{Obergaulinger} M.,  {Aloy} M.~A.,  {Dimmelmeier} H.,   {M{\"u}ller} E.,  2006,
  \aap, 457, 209

\bibitem[\protect\citeauthoryear{{Ott}}{{Ott}}{2009}]{ott:09}
{Ott} C.~D.,  2009, \mn@doi [Class. Quantum Grav.]
  {10.1088/0264-9381/26/6/063001}, \href
  {http://adsabs.harvard.edu/abs/2009CQGra..26f3001O} {26, 063001}

\bibitem[\protect\citeauthoryear{Ott, Burrows, Livne  \& Walder}{Ott
  et~al.}{2004}]{ott:04}
Ott C.~D.,  Burrows A.,  Livne E.,   Walder R.,  2004, \apj, 600, 834

\bibitem[\protect\citeauthoryear{{Ott}, {Burrows}, {Thompson}, {Livne}  \&
  {Walder}}{{Ott} et~al.}{2006}]{ott:06spin}
{Ott} C.~D.,  {Burrows} A.,  {Thompson} T.~A.,  {Livne} E.,   {Walder} R.,
  2006, \apjs, 164, 130

\bibitem[\protect\citeauthoryear{{Ott}, {Dimmelmeier}, {Marek}, {Janka},
  {Hawke}, {Zink}  \& {Schnetter}}{{Ott} et~al.}{2007}]{ott:07prl}
{Ott} C.~D.,  {Dimmelmeier} H.,  {Marek} A.,  {Janka} H.-T.,  {Hawke} I.,
  {Zink} B.,   {Schnetter} E.,  2007, \prl, 98, 261101

\bibitem[\protect\citeauthoryear{{Ott} et~al.,}{{Ott} et~al.}{2012}]{ott:12a}
{Ott} C.~D.,  et~al., 2012, \mn@doi [\prd] {10.1103/PhysRevD.86.024026}, \href
  {http://adsabs.harvard.edu/abs/2012PhRvD..86b4026O} {86, 024026}

\bibitem[\protect\citeauthoryear{{Ott} et~al.,}{{Ott} et~al.}{2013}]{ott:13a}
{Ott} C.~D.,  et~al., 2013, \apj, 768, 115

\bibitem[\protect\citeauthoryear{{Passamonti}, {Haskell}, {Andersson}, {Jones}
  \& {Hawke}}{{Passamonti} et~al.}{2009}]{passamonti:09}
{Passamonti} A.,  {Haskell} B.,  {Andersson} N.,  {Jones} D.~I.,   {Hawke} I.,
  2009, \mn@doi [\mnras] {10.1111/j.1365-2966.2009.14408.x}, \href
  {http://adsabs.harvard.edu/abs/2009MNRAS.394..730P} {394, 730}

\bibitem[\protect\citeauthoryear{{Ruffini} \& {Wheeler}}{{Ruffini} \&
  {Wheeler}}{1971}]{ruffini:71}
{Ruffini} R.,  {Wheeler} J.~A.,  1971, in {Hardy} V.,  {Moore} H.,  eds,
  Proceedings of the Conference on Space Physics, ESRO, Paris, France. p.~45

\bibitem[\protect\citeauthoryear{{Saenz} \& {Shapiro}}{{Saenz} \&
  {Shapiro}}{1978}]{saenzshapiro:78}
{Saenz} R.~A.,  {Shapiro} S.~L.,  1978, \apj, 221, 286

\bibitem[\protect\citeauthoryear{{Scheidegger}, {Fischer}, {Whitehouse}  \&
  {Liebend{\"o}rfer}}{{Scheidegger} et~al.}{2008}]{scheidegger:08}
{Scheidegger} S.,  {Fischer} T.,  {Whitehouse} S.~C.,   {Liebend{\"o}rfer} M.,
  2008, \aap, 490, 231

\bibitem[\protect\citeauthoryear{{Scheidegger}, {Whitehouse}, {K{\"a}ppeli}  \&
  {Liebend{\"o}rfer}}{{Scheidegger} et~al.}{2010}]{scheidegger:10}
{Scheidegger} S.,  {Whitehouse} S.~C.,  {K{\"a}ppeli} R.,   {Liebend{\"o}rfer}
  M.,  2010, \cqg, \href {http://adsabs.harvard.edu/abs/2010CQGra..27k4101S}
  {27, 114101}

\bibitem[\protect\citeauthoryear{{Takiwaki}, {Kotake}  \& {Suwa}}{{Takiwaki}
  et~al.}{2012}]{takiwaki:12}
{Takiwaki} T.,  {Kotake} K.,   {Suwa} Y.,  2012, \mn@doi [\apj]
  {10.1088/0004-637X/749/2/98}, \href
  {http://adsabs.harvard.edu/abs/2012ApJ...749...98T} {749, 98}

\bibitem[\protect\citeauthoryear{{Thorne}}{{Thorne}}{1969a}]{thorne:69a}
{Thorne} K.~S.,  1969a, \apj, 158, 1

\bibitem[\protect\citeauthoryear{{Thorne}}{{Thorne}}{1969b}]{thorne:69b}
{Thorne} K.~S.,  1969b, \apj, 158, 997

\bibitem[\protect\citeauthoryear{{Weber}}{{Weber}}{1966}]{weber:66}
{Weber} J.,  1966, \prl, 17, 1228

\bibitem[\protect\citeauthoryear{{Yamada} \& {Sato}}{{Yamada} \&
  {Sato}}{1995}]{yamadasato:95}
{Yamada} S.,  {Sato} K.,  1995, \apj, 450, 245

\bibitem[\protect\citeauthoryear{{Zwerger} \& {M\"uller}}{{Zwerger} \&
  {M\"uller}}{1997}]{zwerger:97}
{Zwerger} T.,  {M\"uller} E.,  1997, \aap, 320, 209

\makeatother
\end{thebibliography}

\appendix

\section{Oscillation Equations and Boundary Conditions}
\label{equations}

\subsection{Force Exerted by Bounce}

We begin by estimating the magnitude of the force exerted on the fluid during core bounce. The force on a spherical shell is approximately equal to the force required to halt the free fall of the shell. Therefore we require
\beq
\label{intf}
\int dt \ {\rm f} = v_{\rm esc} = \sqrt{\frac{2 GM(r)}{r}}\,\,.
\eeq
Then, using the form of ${\rm f}$ from equation \ref{f2}, we find
\beq
\label{f}
{\rm {\bf  f} } = \sqrt{\frac{2}{\pi}} \ g \ e^{-(t/t_{\rm dyn})^2} {\hat {\bf r} } \,\,.
\eeq

In a centrifugally distorted star, both the magnitude and direction of the force in equation \ref{f} are perturbed (although we will ignore perturbations in the time dependence because its functional form is only approximate), such that the centrifugal perturbation to the force is
\beq
\label{df}
{\rm {\bf  \delta f} } = \sqrt{\frac{2}{\pi}} \ g \ e^{-(t/t_{\rm dyn})^2} \bigg[ \frac{\delta g}{g} {\hat {\bf r} } + \delta {\hat {\bf n} } \bigg] \,\,.
\eeq
Here, $\delta g$ is the perturbation in gravitational acceleration due to the centrifugal distortion, and $\delta {\hat {\bf n} }$ is the perturbation in surface normal to each centrifugally distorted shell. To first order, the centrifugal distortion perturbs the location of each spherical shell, located at radius $r$, by an amount $\delta r = - \epsilon \, r \, Y_{20}$, with $\epsilon \sim (\Omega/\Omega_{\rm dyn})^2$. The perturbation in the gravitational acceleration is approximately $\delta g/g \simeq -2 \delta r/r$, while the perturbation in the surface normal is $\delta {\hat {\bf n} } = - \bnab_\perp \delta r$. Then the perturbed bounce force due to centrifugal distortion is 
\beq
\label{df2}
{\rm {\bf  \delta f} } \simeq A \sqrt{\frac{2}{\pi}} \, \epsilon \, g \, e^{-(t/t_{\rm dyn})^2} \bigg[ 2 \, Y_{20} \, {\hat {\bf r} } + r \bnab_\perp Y_{20} \bigg] \,\,,
\eeq
and $A$ is a constant of order unity that parameterizes the strength of the force.

\subsection{Oscillation Equations}

Next we describe our method of solving the linearized forced oscillation Equations \ref{momeq}-\ref{poisson}. As described in the text, we decompose the perturbed fluid variables into their frequency components. For brevity, we drop the $\omega$ subscript used in the text, but it should be understood that all perturbed variables are the fluid response per unit frequency. We have decomposed the force into spherical harmonics, and do the same for the fluid response, as each spherical harmonic component will couple only to the associated forcing term. The radial component of the Lagrangian displacement is
\beq
\label{xir}
\bxi_r({\bf r}) = U(r) Y_{lm}(\theta,\phi) {\hat{\bf r}}
\eeq
while the horizontal component is
\beq
\label{xip}
\bxi_\perp({\bf r}) = V(r) r \bnab Y_{lm}(\theta,\phi)\,\,.
\eeq
In a rotating star there will also exist a toroidal component to the horizontal displacement, however computation of this term requires the inclusion of Coriolis and centrifugal forces and complicates the procedure. We proceed without including rotational effects, with the understanding that waves, especially at low frequencies, will be significantly altered by the effects of rotation. 

We define the enthalpy perturbation
\beq
\label{psi}
\Psi({\bf r}) = \frac{\delta P({\bf r})}{\rho} + \delta \Phi ({\bf r}) - r \, \delta {\rm f} \, Y_{20} \,\,,
\eeq
where
\beq
\label{fmag}
\delta {\rm f} \equiv \frac{A}{\sqrt{2} \pi} \, \epsilon \, g \, t_{\rm dyn} \, e^{-(\omega t_{\rm dyn}/2)^2} \, .
\eeq
After integrating over angle, the horizontal components of the momentum equation (Equation \ref{momfreq}) yield $V(r) = \Psi(r)/(\omega^2 r)$. 

Upon angular integration, the radial component of Equation \ref{momfreq} becomes
\beq
\label{momrad}
\frac{\partial}{\partial r} \delta P + \frac{g}{c_s^2} \delta P + \rho \delta g + \rho \big( N^2 - \omega^2 \big) U = 2 \, \rho \, \delta {\rm f} \,\,.
\eeq
We have dropped the $(r)$ dependence of the variables for convenience. Transforming to the frequency domain, the continuity equation becomes
\begin{align}
\label{cont2}
\frac{\partial}{\partial r} U &+ \bigg[ \frac{2}{r} - \frac{g}{c_s^2} \bigg] U + \bigg[\frac{1}{c_s^2} - \frac{l(l+1)}{\omega^2 r^2} \bigg] \frac{\delta P}{\rho} - \frac{l(l+1)}{\omega^2 r^2} \delta \Phi \nonumber \\
 &= - \frac{l(l+1)}{\omega^2 r} \delta {\rm f} \,\,.
\end{align}
Finally, Poisson's equation can be written as the two first order differential equations
\beq
\label{dg}
\delta g = \frac{ \partial}{\partial r} \delta \Phi\,\,,
\eeq
\beq
\label{poisson2}
\frac{\partial}{\partial r}\delta g + \frac{2}{r} \delta g - \frac{l(l+1)}{r^2} \delta \Phi - 4 \pi G \rho \bigg[ \frac{1}{\rho c_s^2} \delta P + \frac{N^2}{g} U \bigg] = 0\,\,.
\eeq
It is important to understand that Equations \ref{momrad}-\ref{poisson2} are complex, and each perturbed variable has both a real and imaginary component. The forcing term $\delta {\rm f}$ is purely real and generates differences between the real and imaginary components responsible for energy transport.

To solve Equations \ref{momrad}-\ref{poisson2} we need eight boundary conditions as required for the eight variables composed of the real and imaginary parts of $U$, $\delta P$, $\delta \Phi$, and $\delta g$. The four inner boundary conditions are the usual relations
\beq
\label{in1}
U = \frac{l}{\omega^2 r} \Psi
\eeq
and
\beq
\label{in2}
\delta g = \frac{l}{r} \delta \Phi\,\,.
\eeq

As justified in the text, we require a radiative outer boundary condition. To do this, we use the WKB approximation for the variables at the outer boundary, such that $U \propto e^{i \int k_r dr}$, with $k_r \simeq \pm \omega/c_s$. Since we have defined the time dependence of the waves to be proportional to $e^{-i \omega t}$, the positive value of $k_r$ corresponds to an outgoing wave. In the WKB limit, Equation \ref{cont2} is 
\beq
\label{out1}
i k_r U + \frac{1}{\rho c_s^2} \delta P = 0\,\,,
\eeq
or
\beq
\label{out2}
U = \frac{i}{\omega \rho c_s} \delta P \,\,.
\eeq
Similarly, Equation \ref{poisson2} is approximately
\beq
\label{out3}
\delta g = - \frac{4 \pi i G }{\omega c_s} \delta P\,\,.
\eeq
Equations \ref{out2} and \ref{out3} constitute our four outer boundary conditions.

\section{Wave Dynamics}
\label{dynamics}

\subsection{Work and Energy Flux}

As a check on our numerical calculations and our interpretation of their results, it is helpful to understand the wave energetics. The basic idea is that the forcing term in Equation \ref{cont2} imparts energy into the waves, and this energy is eventually carried away by the outwardly propagating waves. The rate of energy change in the volume interior to radius $r$ due to the waves (in the absence of any forcing) is
\beq
\label{e1}
\dot{E} =  -r^2 \int dA \big(\delta P + \rho \delta \Phi \big) \delta v_r -\int d V \delta \Phi \frac{\partial}{\partial t} \delta \rho\,\,.
\eeq
where $dA$ is a surface area element of the spherical surface at radius $r$. The first term on the right hand side is the energy flux through the surface and the second term is the gravitational work done by the volume. To evaluate this expression, we decompose the perturbations into the frequency domain as in Equation \ref{xifreq}. We then integrate over angle and time to obtain the net energy outflow per unit frequency. The result is
\begin{align}
\label{e2}
\int dt \dot{E}_\omega &=  2 \omega r^2 \big( U_{i,\omega} \Psi_{r,\omega} - U_{r,\omega} \Psi_{i,\omega} \big) \nonumber \\
&+ 2 \omega r^2 \int dr \big( \delta \rho_{r,\omega} \delta \Phi_{i,\omega} - \delta \rho_{i,\omega} \delta \Phi_{r,\omega} \big)\,\,.
\end{align}
Here, the $r$ and $i$ subscripts denote the real and imaginary subscripts of each variable, respectively, and we have reintroduced the $\omega$ subscript to denote the response per unit frequency. The total energy flux (including waves of all frequencies) is found by integrating Equation \ref{e2} over all frequencies. 

We can compare the energy flux of Equation \ref{e2} with the energy deposited into the waves by the forcing term. The work done per unit time by the force is
\beq
\label{eforce}
\dot{E} =  \int dV \rho {\rm {\bf f}} \cdot \delta {\bf v}\,\,. 
\eeq
Once again decomposing the response into its spherical harmonic and frequency components, and integrating over all time, we find the total work done by the force per unit frequency,
\beq
\label{eforce2}
\int dt \dot{E}_\omega =  2 \omega \int dr \rho r^2 {\rm {\bf f}}_\omega \cdot \bxi_{i,\omega} \,\,.
\eeq
Since there are no other wave driving or damping terms, the total work done by the force (Equation \ref{eforce2}) should be identical to the total energy outflow of Equation \ref{e2}, at all radii $r$ and at all angular frequencies $\omega$. To check our numerics we verify that this is indeed the case.

Finally, we can compute the time integrated energy per unit radius per unit frequency. The physical meaning of this quantity is the energy density of a wave packet of angular frequency $\omega$, weighted by the amount of time it spends at any location. The result is 
\beq
\label{edens2}
\int dt \frac{dE_\omega}{dr} =  2 \omega^2 \rho r^2 \big[ |U_\omega|^2 + l(l+1) |V_\omega|^2 \big]\,\,,
\eeq
which is plotted in Figure \ref{NSwavefunction}. The waves spend most of their time in the core, so their time-weighted energy density is largest there. They travel through the outer regions relatively rapidly, without significant reflection, and their energy density is small there. In this case their energy density should be related to the total energy outflow rate via
\beq
\label{ecomp}
\int dt \dot{E}_\omega \simeq \int dt \ c_s \frac{dE_\omega}{dr}\,\,,
\eeq
because $c_s$ is the group velocity of the waves through this region. We verify that Equation \ref{ecomp} is approximately satisfied in the outer layers of our computational domain.

\subsection{Timescales}
\label{wavemode}

It is important to understand the timescales involved for bounce-excited waves in PNSs. The waves of interest have oscillation periods comparable to both the spin period and dynamical timescale of the PNS:
\beq
P \sim P_s \sim P_{\rm dyn} \sim 1 \ {\rm ms}\,\,.
\eeq
In contrast, the PNS and surrounding envelope evolve over longer timescales set by the mass accretion rate,
\beq
P_{\rm evol} \sim 50 \ {\rm ms}\,\,. 
\eeq
The waves have plenty of time to oscillate before they are altered by the evolution of the supernova. However, we should be skeptical of any wave timescales longer than $P_{\rm evol}$, as any processes acting on these timescales will likely be irrelevant compared to the dynamical evolution of the background structure.

To understand timescales associated with the waves, it is helpful to construct a toy problem. We consider the same supernova background structure shown in Figure \ref{NSstruc}. Rather than calculate a forced wave solution as we do in Section \ref{oscillations}, we consider the properties of steady oscillations ocurring at an angular frequency $\omega$. To do this, we solve Equations \ref{momrad}-\ref{poisson2} in the absence of the forcing term. Also, instead of using an outgoing wave outer boundary condition, we set $U=1$ at the outer boundary (this corresponds to a choice of normalization). Physically, this scenario would represent a steady state oscillation due to irradiation by waves with angular frequency $\omega$. The steady state is composed of both ingoing and outgoing waves with equal magnitude so that there is no net energy transport. It resembles an oscillation ``mode" of the background structure, although the mode spectrum is continuous. 

The solution calculated via this technique is the superposition of an ingoing and outgoing wave. Each perturbation variable can thus be expressed in the form
\begin{align}
\label{inout}
\xi_r(r) &= B(r) \bigg[ e^{i \int^r_0 k_r dr} + e^{-i \int^r_0 k_r dr} \bigg] \\
&= B(r) \cos \bigg[\int^r_0 k_r dr\bigg]\,\,,
\end{align}
where $B(r)$ is a wave amplitude and $k_r$ is the radial wave number (Equation \ref{disp}). In the WKB limit this implies
that the wave amplitude is
\beq
\label{Amag}
B = k_r^{-1} \sqrt{ \big(k_r \xi_r \big)^2 + \bigg( \frac{\partial}{\partial r} \xi_r \bigg)^2}\,\,,
\eeq
which is a smooth (non-oscillatory) function of radius and can be calculated from our numerical wave solution. Both the ingoing and outgoing energy flux at any point where the wave is in the WKB limit is
\beq
\label{einout}
\dot{E}(r) \simeq  2 \rho \omega^2 r^2 c_s B^2\,\,.
\eeq
At any point within the star, we can then define a wave leakage timescale
\beq
\label{tleak}
t_{\rm leak}(r) = \frac{E(r)}{\dot{E}(r)}\,\,,
\eeq
where $E(r)$ is the net wave energy contained below the radius $r$
\beq
\label{Er}
E(r) = \int^r_0 \frac{dE}{dr}\,\,,
\eeq
and $dE/dr$ given by Equation \ref{edens2}. 

The quantity $t_{\rm leak}(r)$ is a smoothly varying function that describes the amount of time it would take waves to leak out of a region below radius $r$ in the absence of an ingoing wave flux. Therefore the value of $t_{\rm leak}$ calculated in this toy problem serves as a good proxy for the wave leakage time for our forced oscillation calculations. In the absence of wave reflection within the star, the value of $t_{\rm leak}$ would simply be the wave crossing timescale $t_{\rm cross}$ of Equation \ref{tcross}. However, the wave energy is concentrated within the PNS due to its reflecting edge and so in general $t_{\rm leak}(r) > t_{\rm cross}(r)$ because the wave energy only gradually leaks out of the PNS. 

To estimate the timescale on which waves damp via GW emission, we compute a GW damping timescale
\beq
\label{tgw}
t_{\rm GW}(r) = \frac{E(r)}{\dot{E}_{\rm GW}(r)}\,\,,
\eeq
where $E(r)$ is the wave energy contained within radius $r$, and $\dot{E}_{\rm GW}(r)$ is the GW energy emission rate from Equation \ref{GW2}. The waves become strongly attenuated when their GW damping timescale at a radius $r$ is smaller than the time it takes a wave to propagate past $r$. Except for high frequency waves ($f \gtrsim 2\,{\rm kHz}$), the GW damping timescale is long, $t_{\rm GW}(r) > t_{\rm leak}(r)$. Therefore, GW emission is not likely to be the dominant source of wave damping, and only a small fraction of the energy contained in the fluid motions will be converted to GWs.

\subsection{Wave Amplitudes and Non-linear Effects}
\label{nonlinear}

To formally calculate the wave amplitude at a radial location $r$ and time $t$, one must integrate the response per unit frequency via Equation \ref{xifreq}. If the wave response is sharply peaked at certain values of $\omega$, as it is for frequencies near the PNS oscillation modes, one can approximate the response due to these waves as 
\beq
\label{xiapprox}
\bxi(\omega,t) \sim \Delta \omega \bxi_\omega e^{-i \omega t},
\eeq
where $\Delta \omega$ is the width of the frequency peak in the computed response. This approximation is only valid at times $t \Delta \omega \ll 1$ and at radii corresponding to $t_{\rm cross}(r) \Delta \omega \ll 1$, otherwise waves of different frequencies will deconstructively interfere with one another. 

We use Equation \ref{xiapprox} to estimate the amplitude of waves with frequencies near the PNS quadrupolar f-mode shortly after bounce, as shown in Figure \ref{NSnonlin}. The value of $\xi_r$ is of order kilometers in the inner 100 km of the supernova, translating to displacements of $\sim 20\%$ the radius of the PNS. 

At these large amplitudes our linear calculations begin to break down. We therefore also plot the value of $|k_r \xi_r|$. Modes are strongly non-linear and are expected to quickly dissipate when $|k_r \xi_r| \gtrsim 1$. At larger radii, the waves become increasingly non-linear, which will lead to non-linear wave breaking if the waves make it that far. It is also possible that the waves are dissipated by neutrino damping or turbulent dissipation before they are able to generate non-linear wave breaking. Nonetheless, the fairly strong non-linearity ($|k_r \xi_r| \sim 1)$ of these waves within the PNS indicates that non-linear processes such as three-mode coupling may be important for these waves. One common outcome of such coupling is the transfer of energy to waves with $f \simeq 2 f_{{\rm f-mode}}$. The GW spectra of fully non-linear simulations (K15) contain a peak near twice the f-mode frequency, indicating that non-linear effects may be at play.

\end{document}